\documentclass{jfm}
\usepackage{graphicx}
\usepackage{epstopdf, epsfig}
\usepackage{float}
\usepackage{mwe}
\usepackage{amsmath}
\usepackage[normalem]{ulem}
\usepackage[utf8]{inputenc}
\usepackage[T1]{fontenc}
\usepackage[english]{babel}
\usepackage{color}

\usepackage{bigints}
\usepackage{breqn}
\usepackage{subfigure}
\shorttitle{Viscosity-stratified optimal perturbations}
\shortauthor{R Thakur, A Sharma and R Govindarajan}

\title{Early evolution of optimal perturbations in a viscosity-stratified channel}

\author{Ritabrata Thakur\aff{$*$}\aff{1},
  Arjun Sharma\aff{$*$}\aff{2}
 \and Rama Govindarajan\aff{1}\corresp{\email{rama@icts.res.in}}}

\affiliation{\aff{1} International Centre for Theoretical Sciences, Tata Institute of Fundamental Research, Shivakote, Bengaluru, 560089, India.
\aff{2} Sibley School of Mechanical and Aerospace Engineering, Cornell University, Ithaca, NY, 14853, USA.
\aff{$*$} Equal contribution.}

\begin{document}

\maketitle

\begin{abstract}
This work shows how the early stages of perturbation growth in a viscosity-stratified flow are different from those in a constant-viscosity flow, and how nonlinearity is a crucial ingredient. We derive the viscosity-varying adjoint Navier-Stokes equations, where gradients in viscosity force both the adjoint momentum and the adjoint scalar (here temperature). By the technique of direct-adjoint looping, we obtain the nonlinear optimal perturbation which maximises the perturbation kinetic energy of the nonlinear system. While we study three-dimensional plane Poiseuille (channel) flow with the walls at different temperatures, and a temperature-dependent viscosity, our findings are general for any flow with viscosity variations near walls. The Orr and modified lift-up mechanisms are in operation at low and high perturbation amplitudes respectively at our subcritical Reynolds number. The nonlinear optimal perturbation contains more energy on the hot (less-viscous) side, with a stronger initial lift-up. However, as the flow evolves, the important dynamics shifts to the cold (more-viscous) side, where wide high-speed streaks of low viscosity grow and persist, and strengthen the inflectional quality of the velocity profile. We provide a physical description of this process, and show that the evolution of the linear optimal perturbation misses most of the physics. The Prandtl number does not qualitatively affect the findings at these times. The study of nonlinear optimal perturbations is still in its infancy, and viscosity variations are ubiquitous. We hope that this first work on nonlinear optimal perturbation with viscosity variations will lead to wider studies on transition to turbulence in these flows.

\end{abstract}


%keywords: nonlinear optimisation, viscosity stratification, variational methods, optimal perturbation 

\section{Introduction}

A variation of viscosity in space and time occurs in a vast range of flows. Practically all flows where  composition or temperature are not constant are of varying viscosity. Changes in viscosity are known to affect the stability of the flow dramatically. While an enormous literature is available on viscosity stratification and its effect on linear instability, far less is studied about how it impacts the nonmodal growth of perturbations. Understanding transition to turbulence in shear flow requires understanding how nonmodal perturbations grow and propagate. In recent years, it has been recognised \citep{pringle2010using, cherubini2010rapid, pringle2012minimal} that studying the nonlinear optimal perturbations is essential to this effort. The present study is the first to our knowledge on nonlinear optimal perturbations in viscosity-stratified flows. Our interest is in a gentle variation of viscosity rather than a sharp one, and we choose a pressure-driven channel flow with the walls maintained at different temperatures as a prototypical model flow to reveal the essential physics. Further, we are interested in short term optimisation, to underline how viscosity varying flows already depart considerably from constant viscosity flows. We set gravity to zero in this study to isolate the effects of viscosity variation.

The interaction of viscosity stratification and shear can lead to both suppression and enhancement of flow instabilities (for a review, see \citet{govindarajan2014instabilities}). A viscosity jump across an interface can give rise to linear instability at any Reynolds number (see e.g. \citet{yih1967instability}). On the other hand, a lowering of viscosity near a wall has been studied for decades as a means to stabilise shear flow and to thus achieve drag reduction, e.g., in lubricating oil pipelines  \citep{preziosi1989lubricated}. Composition variation and the introduction of polymers, whence besides elasticity, viscosity stratification resulting due to shear thinning can be important, have been explored over the years. In aerospace applications \citep{mack1984boundary}, a viscosity reduction near the wall in a boundary layer can provide a fuller and more stable velocity profile. By virtue of viscosity (see e.g. \cite{schmid2002stability}) and its spatial gradients \citep{govindarajan2004effect} being multiplied by the highest derivatives in the stability equations, we are presented with a singular perturbation problem. In other words, however high the Reynolds number (however small the viscosity), viscosity and its variations can have a large effect on the flow. For example, \cite{ranganathan2001stabilization} showed that a ten percent change in viscosity across a thin layer can, if overlapped with the critical layer of the least stable eigenmode, give rise to an order of magnitude change in the critical Reynolds number $Re_c$ of $5772.2$ in a channel. The effect of wall heating and subsequent viscosity changes on a fully developed turbulent flow has been studied using direct numerical simulations (DNS) for both a boundary layer \citep{lee2013effect} and a channel flow \citep{zonta2012modulation}. \cite{zonta2012modulation} find vortical structures to be more populated near the colder (more viscous) wall as compared to the hotter (less viscous) wall, while \citet{lee2013effect} find that vortical structures near the heated wall are unaffected, whereas away from the wall, they become sparser with wall heating. The effects of a continuous variation of viscosity have also been investigated in the linear stability studies of \citet{potter1972stability, schafer1993stability, wall1996linear, sameen2007effect}. 

For a channel flow below $Re_c$, a traditional normal-mode analysis predicts that the energy of no single eigenmode can grow in isolation. However, the linear stability operator of the flow, obtained by linearising the Navier-Stokes equations about a laminar flow and posing the resulting Orr-Sommerfeld and Squire equations as an eigenvalue problem, is non-normal. Hence, a transient (algebraic) growth in energy can occur in the flow due to the superimposition of suitably arranged eigenmodes at intermediate time \citep{reddy1993energy,trefethen1993hydrodynamic}. Shear makes the governing operator non-normal and transient growth is experienced, for example, in channel flow, pipe flow and Couette flow. If the transient growth is large enough, nonlinear mechanisms could be activated.  It is now accepted that findings such as that the transition to turbulence in a channel at $\Rey\approx1000$ (e.g., \citet{orszag1980subcritical, carlson1982flow}) is a manifestation of these nonmodal and nonlinear mechanisms. For such flows, non-modal analyses complement modal analysis in fully understanding the behaviour \citep{boberg1988onset, butler1992three, trefethen1993hydrodynamic, trefethen2005spectra} (for a review, see \citet{schmid2007nonmodal}). A linear non-modal study often optimises for the energy growth of an infinitesimal initial perturbation over all possible initial conditions, and from the singular value decomposition (SVD) of the linear operator, reveals the optimal perturbation, i.e., the initial perturbation that leads to the largest transient growth in the linear regime.

%,  and the DNS of \citet{tuckerman2014turbulent}
%At finite perturbation energies, nonlinear mechanisms are important and an extension of the nonmodal 
For any amplitude of initial perturbation, the optimal perturbation can be obtained by an adjoint-based iterative optimisation procedure with the full, or linearised, Navier-Stokes equations as in \citet{schmid2007nonmodal}. This procedure involves repeated computations of adjoint fields and the sensitivity of a cost functional to changes in the initial perturbation. It has been applied to the  Navier-Stokes equations for control of fluid flow by \citet{abergel1990some, bewley2000general, zuccher2004algebraic} among others, and to numerically calculate the optimal perturbations and the associated transient growth, within the framework of the linearised as well as of the nonlinear Navier-Stokes equations, as in \citet{monokrousos2011nonequilibrium}, \citet{foures2013localization}, \citet{kaminski2014transient}, \citet{marcotte2018optimal}, \citet{vermach2018optimal} (for a review, see \citet{kerswell2018nonlinear}). The nonlinear optimal perturbation has been found to have a different spatial structure from the linear optimal perturbation \citep{rabin2012triggering}. Since the full nonlinear equations are optimised, the nonlinear optimal perturbation leads to a larger transient growth \citep{cherubini2010rapid, pringle2010using,  luchini2014adjoint}. This hints at the importance of the nonlinearity in the nonmodal analysis, and indicates that the search for a minimal seed for turbulence onset must involve studying the time evolution of the nonlinear optimal (see e.g. \citet{pringle2012minimal}).

In this paper, we investigate the sole effects viscosity stratification on the optimal perturbation and the resultant transient growth at early times. Our central idea is to investigate how the process of subcritical perturbation growth in the nonlinear regime is affected by viscosity variations. We consider the full nonlinear Navier-Stokes equations, modified to account for varying viscosity, and derive the adjoint viscosity-stratified Navier Stokes equations. We then formulate a nonlinear stability theory using the adjoint-based optimisation technique. We utilise this framework to calculate the optimal perturbation for a certain fixed target-time. Performing studies with very small and more significant initial perturbation amplitudes, our findings show how nonlinearity is a crucial part of the initial evolution, although the Orr and lift-up mechanisms in operation have linear underpinnings. The evolution of initial perturbation which maximises linear energy growth is restricted to the hot wall, whereas optimising for nonlinear energy growth shows how the cold wall is more important, with persistent streaks and velocity profiles becoming increasingly inflectional.

\section{Governing equations and problem formulation}

%A constant pressure gradient directs the mean flow towards the positive $x$ direction. 

\begin{figure}
\centering
\centerline{\includegraphics[width=0.8\textwidth, height=0.8\textheight,keepaspectratio]{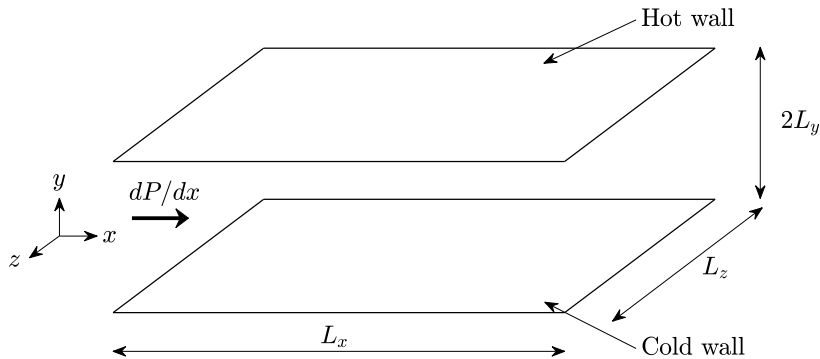}}
    \caption{The flow domain being studied. The flow is from left to right, driven by the mean pressure gradient $dP/dx$. $L_x = 2\upi L_y$ is the streamwise length, $L_z = \upi L_y$ is the spanwise length, and $L_y$ is the half-width of the channel. The hot and the cold walls at $y = \pm L_y$ are kept at constant but different temperatures.}
\label{plot:channel}
\end{figure}
%\textcolor{red}{(possibly) 

\begin{figure}
\centering
\centerline{\includegraphics[width=0.95\textwidth, height=0.95\textheight,keepaspectratio]{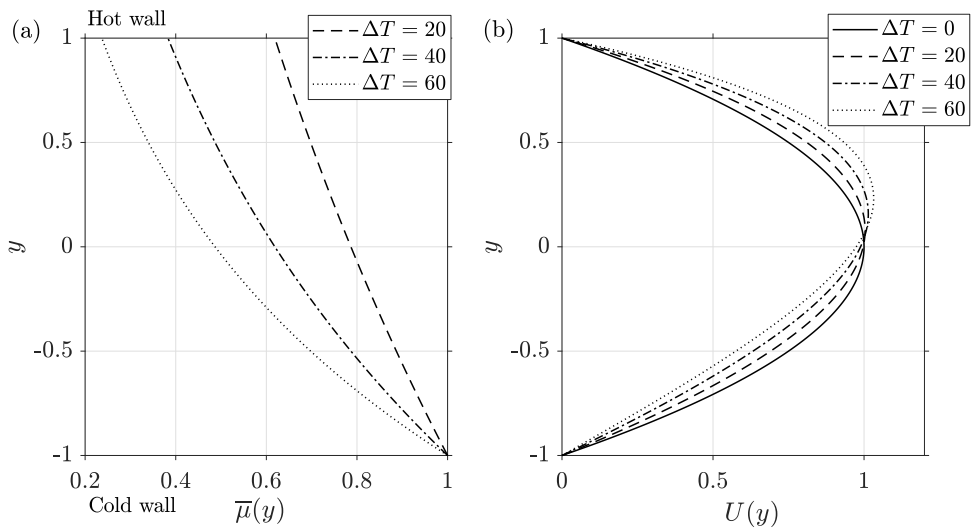}}
    \caption{The wall-normal ($y$) profiles, for various temperature differences $\Delta T$ between the walls, of (a) base viscosity $\overline{\mu}(y)$ as given by equation \eqref{eq:visc_model}. The profile for $\Delta T = 0$ is a vertical line at $\mu(y) = 1$. The ratios of viscosity between the top (hot) and the bottom (cold) wall are $0.61$ for $\Delta T = 20$ K (dashed line), $0.38$ for $\Delta T = 40$ K (dash-dotted line), and $0.23$ for $\Delta T = 60 K$ (dotted line). (b) The unperturbed streamwise laminar velocity $U(y)$, normalised to have equal volumetric flux through the channel for unstratified case (solid line) and different $\Delta T$.}
\label{plot:profiles}
\end{figure}

We study pressure-driven flow through a three-dimensional channel bounded by two parallel walls, kept fixed at $y = \pm L_y$ as depicted in figure \ref{plot:channel}. The mean pressure gradient $dP/dx$ forces the flow in the $x$ direction. Hence, $x$ is the streamwise direction and $z$ the spanwise direction. The temperature of both walls is kept constant, with the wall at $y = L_y$  at a higher temperature than the wall at $y = -L_y$. There is no gravity in this problem, and non-Boussinesq effects arising from density change due to temperature variations are neglected. The half-width, $L_y$, of the channel is chosen as our length scale. The nondimensional size of the channel is fixed at $2\upi, 2, \upi$ in the $x$, $y$, and $z$ directions, respectively. 

The unperturbed laminar flow through the channel is our base state. Three-dimensional perturbations are introduced over this base state.
%and their adjoints by implicit-explicit time marching scheme.
%\subsection{Numerical method}
The nondimensional governing equations for a viscosity-stratified flow read as,
\begin{dmath}
\frac{\partial u_i}{\partial x_i}=0,\label{eq:Direct_incompressibility}
\end{dmath}
\begin{dmath}
\frac{\partial{u_i}}{\partial{t}}+(U_j+u_j)\frac{\partial u_i}{\partial x_j} +u_j\frac{\partial U_i}{\partial x_j}=-\frac{\partial p}{\partial x_i}+{\frac{2 \beta}{\Rey}}{\frac{\partial}{\partial x_j}}\left[{\mu}\big(s_{ij}+S_{ij}\big)+\bar{\mu}s_{ij}\right],\label{eq:Direct_momentum}
\end{dmath}
\begin{dmath}
\frac{\partial{T}}{\partial{t}}+(U_j+u_j)\frac{\partial T}{\partial x_j} +u_j\frac{\partial(\overline{T}+T_0)}{\partial x_j}=\frac{1}{{\Rey }\Pran} \frac{\partial^2T}{\partial x_j^2}.\label{eq:Direct_scalar}
\end{dmath}
Here $U_j =\delta_{j1} U(y)$ is the laminar base state, consisting only of a  streamwise component, $u_j(x,y,z,t)$ are the components of the perturbation velocity $\boldsymbol{u}(\boldsymbol{x},t)$ and $p(\boldsymbol{x},t)$ is the perturbation pressure. $x, \ y$, and $z$ are referred to as $x_1$, $x_2$, and $x_3$, respectively. 
%The viscosity ratio $\beta$ is defined in equation \eqref{viscrat}.
\begin{equation}
    S_{ij} = \frac{1}{2} \left( \frac{\partial U_i}{\partial x_j}+\frac{\partial U_j}{\partial x_i} \right) \quad \textrm{and} \quad s_{ij} = \frac{1}{2} \left( \frac{\partial u_i}{\partial x_j}+\frac{\partial u_j}{\partial x_i} \right)
\end{equation}
are the base and the perturbation velocity strain tensors. $T(\boldsymbol{x},t)$ is the perturbation temperature, and the base state temperature $\overline T(y)+T_0$ is linear in $y$, varying from the reference temperature $T_0$ at the bottom wall to  $T_0 + \Delta T$ at the top wall. The base and perturbation  viscosities, $\overline{\mu}(T)$ and $\mu(T)$ respectively, are functions of temperature alone, and are defined in section \ref{subsec:base_state}. The Reynolds number $\Rey$ and the viscosity ratio $\beta$ are defined in section \ref{subsec:scaling_re}. $\Pran= \mu_{0} c_p/\rho k$ is the Prandtl number, where $\mu_{0}$ is the viscosity at the reference temperature $T_0$, $c_p$ the specific heat at constant pressure, and $k$ the thermal conductivity of the fluid. The density $\rho$ of the fluid is taken to be a constant. 

The initial condition for perturbation velocity
\begin{equation}\label{eq:vel_init}
    \boldsymbol{u}(\boldsymbol{x},0) = \boldsymbol{u_0}(\boldsymbol{x}),
\end{equation}
is usually a random noise and temperature perturbations $T(\boldsymbol{x},0)$ are initialised to zero. Barring the mean pressure drop $dP/dx$ which is linear, all variables of the flow are prescribed to be periodic at the domain boundaries in $x$ and $z$. No-slip velocity boundary conditions are imposed at the walls. 

%The difference in the temperature between the walls,
%\begin{equation}
%\Delta T \equiv T(y=1) - T(y=-1)  \geq 0.    
%\end{equation}

We will refer to equation \eqref{eq:Direct_momentum} as the modified Navier-Stokes equation, valid for viscosity-stratified flow. The set of equations \eqref{eq:Direct_incompressibility}-\eqref{eq:Direct_scalar} are referred to as the `direct' equations to distinguish them from another set of equations called the `adjoint' equations, to be introduced in section \ref{dal}. The variables appearing in equations \eqref{eq:Direct_incompressibility}-\eqref{eq:Direct_scalar} will be called direct variables.

\subsection{Viscosity model and the base state}\label{subsec:base_state}

The local non-dimensional viscosity $\mu_{tot}$ in the flow is modelled as an exponential function of the total temperature $T_{tot} = \overline T(y)+T_0 + T$, following \citet{wall1996linear}, as
\begin{equation}
    \mu_{tot} \equiv \overline{\mu} + \mu = \frac{\exp(-\kappa T_{tot})}{\exp(-\kappa T_0)}, \quad {\rm where} \quad  \overline{\mu}=\frac{\exp[-\kappa (\overline{T}(y)+T_0)]}{\exp(-\kappa T_0)}.
    \label{eq:visc_model}
\end{equation}
The viscosity of the cold wall is used as the scale here. With the constant $\kappa$ chosen to be 0.012 per degree Kelvin, this function closely follows the viscosity of water in our temperature range. Since the density of water varies by less than 2 parts in a 1000 for the largest temperature difference, variations in kinematic viscosity are mainly from changes in dynamic viscosity. As is typical of liquids, the viscosity decreases with an increase in temperature, as shown in figure \ref{plot:profiles}(a). The laminar base profile of the streamwise velocity given by \citep{wall1996linear} 
\begin{equation}
    U(y) = \frac{-2 \alpha}{\kappa\Delta{T}}\left[1 + \coth \kappa\Delta{T} + (y - \coth \kappa\Delta{T})\exp(\kappa\Delta{T}(1+y)) \right],
\end{equation}
where
\begin{equation}
\alpha=\frac{2\kappa\Delta T}{3}\frac{1}{-2(1+\coth{\kappa\Delta T})+(\exp(2\kappa\Delta T)-1)/(\kappa\Delta T)^3},
\end{equation}
allows for the same non-dimensional volumetric flow rate through the channel for different $\Delta T$ as shown in figure \ref{plot:profiles}(b). The nondimensional mean pressure gradient is
\begin{equation}\label{eq:meanpressure}
    \frac{d P}{dx}=-\frac{2\alpha\beta}{\Rey}.
\end{equation}

%We study four different temperature differences $\Delta T=$ 0 (unstratified), 20 K, 40 K and 60 K. 
\subsection{The Reynolds number}\label{subsec:scaling_re}
 %\sout{perturbation energy growths} \textcolor{red}{
In order to make a fair comparison between the growth of perturbation energy in a stratified flow and an unstratified flow, a careful definition of the Reynolds number is required. As the laminar base velocity profile in a stratified channel is asymmetric about $y=0$ (figure \ref{plot:profiles}(b)), the centerline velocity is not a standard velocity scale across different stratification levels, whereas the volume flux is. Secondly, the viscosity in the channel decreases continuously when moving away from the cold wall at $y=-1$ (figure \ref{plot:profiles}(a)). If, for example, $\Rey$ was defined based only on the viscosity at the cold wall, then the effective Reynolds number of the stratified channel would be higher than this value, and consequently the perturbation energy growth could be expected to be higher. So, we choose the space-averaged mean viscosity as our viscosity scale to define $\Rey$. The Reynolds number used in this paper is

%(RT to AS: \textbf{I have removed the $\rho$ multiplied in the numerator as the $\mu$ here is kinematic rather than dynamic viscosity. plz check once. AS: I don't think it should be removed. I think $\mu$ is the dynamic viscosity. But it does not matter too much since we are not changing density and hence we can call it whatever we want. I think generally people use $\mu$ for dynamic and $\nu$ for kinematic viscosity} - \textcolor{red}{ we are using $\mu$ in eq 2.2 which is kinematic viscosity and hence we have to maintain consistency},

\begin{equation}
\Rey \equiv \rho L_y \frac{\int\limits_{-L_y}^{L_y} 1.5 U(y) dy}{\int\limits_{-L_y}^{L_y} \overline{\mu}_d dy} \quad = \frac{1.5 \rho L_y \langle U\rangle}{\langle\overline{\mu}_d\rangle},
\end{equation}
where $\overline{\mu}_d(T)$ is the dimensional base viscosity of the fluid, and the angle brackets represent an average in the wall normal direction $y$. A factor of 1.5 is incorporated for ease of comparison with earlier studies on unstratified flow which use the centerline velocity as the velocity scale. The dimensional viscosity must therefore be scaled by the average viscosity in the channel, but for ease of comparison, we have scaled it by its value at the cold wall. This is adjusted for, by the introduction in equations \eqref{eq:Direct_momentum} and \eqref{eq:meanpressure} of the viscosity ratio

\begin{equation}
\beta = \frac{\overline{\mu}_d(T_0)}{\langle\overline{\mu}_d\rangle}.
\label{viscrat}
\end{equation}
%which is a constant for a given $\Delta T$ \textcolor{red}{and appears in the mean pressure gradient in equation . 
In this paper, we remain in the subcritical regime by fixing $\Rey$ at 500, which also allows for validation against unstratified channel flow result of \citet{vermach2018optimal}.

\subsection{Nonmodal analysis using `direct-adjoint looping'}\label{dal}

The nonlinear nonmodal analysis is formulated in terms of an optimisation procedure termed `direct-adjoint looping', to find the largest nonmodal energy growth and the optimal perturbation of the flow that causes this growth \citep{arratia2013transient, foures2014optimal, kaminski2014transient, vermach2018optimal}. To effect this, we need to define a cost functional which includes some measure of energy, and the aim of the optimisation procedure would be to maximise this cost functional. 
%The procedure can obtain the sensitivity of initial perturbation $u$ in \eqref{eq:Direct_momentum} to the nonmodal energy growth and can find the initial perturbation $u$ which will optimise a given cost functional. 
Especially when density or viscosity or any flow component varies with space and time, there are many choices that may be made for the cost functional, and each choice could lead to a different optimal perturbation.  For example, \citet{foures2014optimal} show interestingly that energy optimization leads to weak mixing, but optimal perturbations obtained from mixing optimization are very effective in mixing, though evolving to lower energies. Thus the aims of each study are critical in choosing an appropriate cost functional.
%The level of mixing is a quantity of importance in such flows, and recent studies \citep{vermach2018optimal,marcotte2018optimal} have found that the perturbation giving the fastest growing energy need not be the one which gives the best mixing. 

%To investigate nonmodal energy growth, a measure can be the growth of the time-integrated perturbation kinetic energy, defined as,
In this first attempt to understand the optimal perturbations in a viscosity-stratified channel flow, we study the growth of kinetic energy of the velocity perturbations. As noted in previous studies \citep{foures2014optimal,vermach2018optimal} perturbations growing through a given time horizon may not have largest energy precisely at a target-time. To account for this, we choose the ratio of the integral over time, up to a preset target-time, of the perturbation kinetic energy, to the initial perturbation kinetic energy, as our cost functional. The time-integrated perturbation kinetic energy of the flow is defined as
%Mathematically, this measure of growth can be expressed as,
\begin{dmath}\label{gteqn}
    G(t) = \frac{\gamma}{2} \int_{0}^{\mathcal{T}} ||\boldsymbol{u}(\boldsymbol{x},t)||_{\mathcal{V}}^{2} dt,
\end{dmath}
where $||\boldsymbol{u}(\boldsymbol{x},t)||_{\mathcal{V}}$ is the total (integrated over the channel volume $\mathcal{V}$) $L^2$-norm of the velocity perturbations $\boldsymbol{u}(\boldsymbol{x},t)$. Note that the math-calligraphy symbol $\mathcal{T}$ for the target-time is distinguished from the italics $T$ for temperature. $\gamma$ is a constant with units of inverse time, and has been set to unity throughout this study. $\mathcal{T}$ is non-dimensionalised with the advective time scale, i.e., $L_y/1.5\langle U \rangle$, a constant across the various stratification levels studied here. Time-integration includes effects from the intermediate-time dynamics of the flow as opposed to just the energy at the target-time $\mathcal{T}$. The other quantity needed to construct the cost functional is the total initial perturbation kinetic energy
\begin{dmath}\label{e0eqn}
    E_0 = \frac{1}{2} ||\boldsymbol{u_0}(\boldsymbol{x})||_{\mathcal{V}}^{2}.
\end{dmath}
The cost functional $\mathcal{J}(t)$ of our interest is
\begin{equation}\label{cost_function}
    \mathcal{J}(t) = \frac{G(t)}{E_{0}}.
\end{equation}

%Unlike in a linear adjoint problem, the amplitude  $E_0$ of the initial energy determines the evolution, and therefore the nonlinear optimal will depend on $E_0$.

%The quantity in equation \eqref{gteqn} has also been optimised for in a 2D channel in \citet{foures2014optimal} and in a 3D channel in \citet{vermach2018optimal}, and hence using this form helps us maintain a standard. 

% concept in the formulation of this nonomodal stability theory is the initial energy  of the perturbation field. 

%For nonmodal analysis to calculate the optimal perturbation and the resultant transient growth, we need to define the framework of the 

\noindent Our aim is to find the optimal perturbation $\boldsymbol{u_0}(\boldsymbol{x},0)_{opt}$ to get
\begin{equation}\label{eq:jopt}
 \mathcal{J}_{opt}(t) \equiv \mathcal{J}_{max}(t) = \frac{G_{opt}(t)}{E_{0}},
\end{equation}
with a fixed inital energy $E_0$.

%As we fix the initial energy, we have $||\boldsymbol{u_0}(\boldsymbol{x},0)_{opt}||_{\mathcal{V}}^{2}$ = $||\boldsymbol{u_0}(\boldsymbol{x},0)||_{\mathcal{V}}^{2}.$

% (It is to be mentioned that as we are interested in the growth of energy, our optimal condition is the maximum value. But for certain other problems where the interest could be to reduce a certain quantity or norm, e.g. turbulent dissipation, a modified $\mathcal{J}_{opt}(\mathcal{T})$ would actually be $\mathcal{J}_{min}(\mathcal{T})$.) 

To formulate the optimisation procedure, we first define a Lagrangian $\mathcal{L}$ which is the cost functional $\mathcal{J}(t)$ in equation \eqref{cost_function}, constrained by the incompressibility condition \eqref{eq:Direct_incompressibility}, the modified viscosity-stratified Navier-Stokes equations \eqref{eq:Direct_momentum}, the temperature equation \eqref{eq:Direct_scalar}, and the initial velocity conditions of the flow \eqref{eq:vel_init}.
%  We can impose the initial energy by using a Lagrange multiplier (as can be seen in the last term of Eq \ref{bigeq}). 
The constrained Lagrangian $\mathcal{L}$ is
\begin{dmath}\label{bigeq}
\mathcal{L}=\mathcal{J}(t)-\Bigg[\frac{\partial{u_i}}{\partial{t}}+(U_j+u_j)\frac{\partial u_i}{\partial x_j} +u_j\frac{\partial U_i}{\partial x_j}+\frac{\partial p}{\partial x_i}-{\frac{2\beta}{\Rey}}{\frac{\partial}{\partial x_j}}\Big({\mu}\big(s_{ij}+S_{ij}\big)+\bar{\mu}s_{ij}\Big),v_i\Bigg]-\Bigg[\frac{\partial{T}}{\partial{t}}+(U_j+u_j)\frac{\partial T}{\partial x_j} +u_j\frac{\partial(\overline{T}+T_0)}{\partial x_j}-\frac{1}{{\Rey }\Pran} \frac{\partial^2T}{\partial x_j^2},\tau\Bigg]-\Bigg[\frac{\partial u_i}{\partial x_i},q\Bigg]-\langle\langle u_i(0)-u_{0,i},v_{0,i}\rangle\rangle,
\end{dmath}
where $\langle\langle a \rangle\rangle$ is the volume integral of some quantity $a$ %\rg{the single angle bracket is a y average, so we need double angle bracket here, remove this if you agree},
%\begin{dmath}
%\langle a \rangle = \int\!\!\!\int\!\!\!\int a %d\mathcal{V},
%\end{dmath}
and $[a,b]$ is the inner product of some quantities $a$ and $b$. $u_i(0) = u_{0,i}$ are the components of the initial perturbation velocity $\boldsymbol{u_0}(\boldsymbol{x})$. The new quantities $v_i$, $q$, $\tau$, and $v_{0,i}$ are the adjoint velocity, adjoint pressure, adjoint temperature, and adjoint velocity initial condition corresponding to direct variables $u_i$, $p$, $T$, and $u_{0,i}$.

The variation of $\mathcal{L}$ with respect to all the variables and their corresponding adjoints are independent of each other. At the maximum of the cost functional $\mathcal{J}(t)$ and hence $\mathcal{L}$ in equation \eqref{bigeq}, the variational derivatives identically vanish. The vanishing variational derivatives with respect to $p$, $u_i$, and $T$ give us the adjoint (continuity, momentum, and temperature) equations governing the time evolution of adjoint variables, $v_i$, $q$, and $\tau$ and the required `initial' conditions at $t=\mathcal{T}$,
\begin{dmath}
\frac{\partial v_i}{\partial x_i}=0,\label{eq:Adjoint_incompressibility}
\end{dmath}
\begin{dmath}
\frac{\partial v_i}{\partial t}- v_j\frac{\partial (u_j+U_j)}{\partial x_i}+\frac{\partial (v_i(U_j+u_j))}{\partial x_j}+\frac{\beta}{\Rey}\frac{\partial}{\partial x_j}\Bigg[({\mu}+\overline{\mu})\Big(\frac{\partial v_i}{\partial x_j}+\frac{\partial v_j}{\partial x_i}\Big)\Bigg]-\tau\frac{\partial (T+\overline{T}+T_0)}{\partial x_i}+\frac{\partial q}{\partial x_i}+\gamma u_i=0,\label{adjoint_mom}
\end{dmath}
\begin{dmath}
\frac{\partial \tau}{\partial t}+(U_j+u_j)\frac{\partial \tau}{\partial x_j}+\frac{1}{\Rey\Pran}\frac{\partial^2\tau}{\partial x_j^2}-\frac{2\beta}{\Rey}\Bigg[\frac{\partial\mu}{\partial T}\big(s_{ij}+S_{ij}\big)+\frac{\partial\bar{\mu}}{\partial T}s_{ij}\Bigg]\frac{\partial v_i}{\partial x_j}=0,\label{eq:Adjoint_scalar}
\end{dmath}
\begin{equation}\label{eq:Adjoint_IC}
v_i(\mathcal{T})=0, \: \: \tau(\mathcal{T})=0.
\end{equation}
% These viscosity-stratified adjoint equations are derived for the first time to the best of our knowledge. 

%\citet{kaminski2017nonlinear} \sout{for linear perturbations, once the non-linear adjoint-real terms, i.e., $\mathcal{O}(u_iv_j)$, $\mathcal{O}(u_i\tau)$ in the above equations and gravity (Richardson number, $Ri$) in equations of} \citet{kaminski2017nonlinear} \sout{are ignored}

Similarly, the derivatives of $\mathcal{L}$ in equation \eqref{bigeq} with respect to the adjoint variables, give us back the direct equations  \eqref{eq:Direct_incompressibility}-\eqref{eq:Direct_scalar}. So, $v_i$, $q$, and $\tau$ help impose the direct equations as constraints on $\mathcal{L}$ and hence are Lagrange multipliers. Equations \eqref{eq:Adjoint_incompressibility}-\eqref{eq:Adjoint_scalar} are the adjoint equations corresponding to the direct equations \eqref{eq:Direct_incompressibility}-\eqref{eq:Direct_scalar}. For a constant viscosity flow, these adjoint equations reduce to those derived by \citet{vermach2018optimal} for mixing of a passive scalar. $v_i$ and $q$ have the same dimensions as the direct variables $u_i$ and $p$. But $\tau$ behaves as the square of a velocity per unit temperature. Nevertheless, we refer to it as adjoint temperature since its evolution equation \eqref{eq:Adjoint_scalar} is derived by taking a variation of $\mathcal{L}$ in equation \eqref{bigeq} with respect to $T$. We notice that in the absence of viscosity stratification (the last term in equation \eqref{eq:Adjoint_scalar} with the coefficient of $2\beta/\Rey$, vanishes), and since we have no gravity, the solution to equation \eqref{eq:Adjoint_scalar} is just $\tau=0$, and the temperature term will drop out of the adjoint momentum equation \eqref{adjoint_mom}. The sign of the diffusion of adjoint momentum and temperature in equations \eqref{adjoint_mom} and \eqref{eq:Adjoint_scalar} imply that only during backward time evolution, i.e., from $t=\mathcal{T}$ to 0, these equations are well posed. With this formulation, the adjoint variables at $t=0$ give information on the gradient taking us towards the optimal perturbation. Another constraint that is missing in equation \eqref{bigeq} is the imposition of a fixed $E_0$. This could have been done with a Lagrange multiplier in equation \eqref{bigeq}, but that has been found to be numerically expensive and delicate \citep{foures2013localization}. Hence, during the update of the initial perturbation, carried out to march towards the optimal perturbation based on the gradient information, we use a rotation technique to constrain the $E_0$ of the updated perturbations on a fixed energy hypersphere, as described in detail in \citet{foures2013localization}. The adjoint equations for a viscosity-stratified flow are derived here for the first time to our knowledge. We see new terms involving gradients in viscosity, both of the mean and of the perturbations, entering the adjoint velocity as well as the adjoint temperature equations.

The optimisation procedure consists of an iteration in time. We start with a guess of the optimal perturbation $\boldsymbol{u_0}(\boldsymbol{x},0)_{opt}$, which is usually a random noise. The direct equations \eqref{eq:Direct_incompressibility}-\eqref{eq:Direct_scalar} are marched forward in time till $t=\mathcal{T}$, where the adjoint variables are initialised according to equation \eqref{eq:Adjoint_IC}. The adjoint equations \eqref{eq:Adjoint_incompressibility}-\eqref{eq:Adjoint_scalar} are marched backwards in time from $t=\mathcal{T}$ to 0 where we update our initial guess $\boldsymbol{u_0}(\boldsymbol{x},0)$ in the gradient direction given by the adjoint variables. It is to be noted that the adjoint equations have terms which are products of direct and adjoint variables. So, the direct variables have to be stored at each time step when the direct equations are being solved, to be used in the adjoint equations while marching backward in time. To find the optimal perturbation within a set numerical tolerance, we have to iterate repeatedly and gradually march according to the gradient information and monitor a residual, as defined in other studies like \citet{vermach2018optimal}, which denotes whether we have converged to the actual optimal perturbation. For all the cases studied, when the rotation technique of \cite{foures2013localization} converges (as discussed in appendix A), we find the residual to be $O(10^{-3}-10^{-4})$, and we decree the optimiser to have found the optimal perturbation. This optimisation procedure has been termed direct-adjoint looping. 

% \sout{at this point}
%\sout{The rotation-based method of} \cite{foures2013localization} \sout{converges when the required update angle in \rg{$E_0$ falls below a critical small value, i.e., practically no update} is required in the cost functional.} 

%\sout{in the direction of the optimal perturbation} 
%Hence, starting from a random noise $\boldsymbol{u_0}(\boldsymbol{x},0)$, we arrive at some approximation of the optimal $\boldsymbol{u_0}(\boldsymbol{x},0)_{opt}$.

Whether or not nonlinear mechanisms will be important in the evolution will depend on $E_0$. With $E_{0} = 10^{-2}$, as used in nonlinear optimisation studies of \citet{cherubini2010rapid, foures2014optimal, vermach2018optimal}, the perturbations are at most an order of magnitude smaller than the laminar base flow, and will hence trigger nonlinear mechanisms. On the other hand, with a small $E_0$ of $O(10^{-8})$, nonlinear mechanisms remain unimportant throughout our time horizon, perturbations being several orders of magnitude smaller than the laminar base flow, and their products vanishingly small. For the highest $E_{0}$ of $10^{-2}$ the grid in our study is set at 100 $\times$ 209 $\times$ 50 points in the $x$, $y$, and $z$ directions and we validate our solver with \citet{vermach2018optimal}. More details on the numerical method are given in appendix A. Unless otherwise specified, we set a Prandtl number $\Pran=7$ in our simulations. The target-time of optimisation is fixed at $\mathcal{T}=4$ and we study the linear and nonlinear optimal perturbations and the mechanism behind their evolution, for an unstratified flow, and for stratified flows with temperature differences between the upper and the lower channel walls at $\Delta T = 20$ K, $40$ K, and $60$ K. \citet{kaminski2017nonlinear} studied the non-linear evolution of the linear optimal perturbations in a density-stratified flow and found the linear optimal perturbations to be sufficient to trigger non-linear effects when evolved with sufficiently large $E_0$.
%using full non-linear direct equations. 
However, as we will show later for viscosity-stratified flows, the linear optimal perturbation is qualitatively different in structure from the non-linear optimal perturbation, and hence leads to qualitative and quantitative differences even when scaled to have large $E_0$. For unstratified flow despite the clear structural differences in the linear and non-linear optimal perturbations, we find similar qualitative and quantitative evolution of the two, when initialised at $E_0=10^{-2}$. Thus, the non-linear $O(u_iv_j)$, $O(u_i\tau)$ terms in the adjoint equations \eqref{adjoint_mom} and \eqref{eq:Adjoint_scalar} are critical, especially for the viscosity-stratified flow. We also remark briefly upon the effect of Prandtl number on the evolution of the non-linear optimal perturbation.

\section{Viscosity-stratified optimal perturbations and their evolution}

\subsection {The linear optimal and its evolution}
\begin{figure}
    \centering
    \centerline{\includegraphics[width=1.0\textwidth, height=1.0\textheight,keepaspectratio]{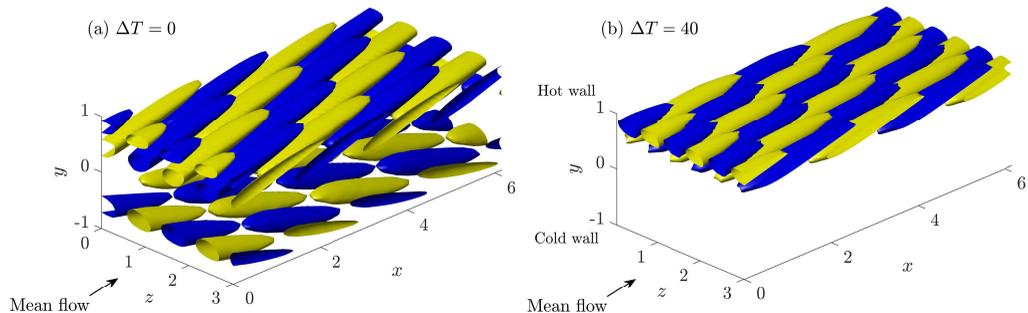}}
    \caption{Three dimensional linear optimal perturbation ($E_0=10^{-8}$), which maximises the cost functional in equation \eqref{cost_function} for (a) unstratified ($\Delta T=0$) and (b) stratified ($\Delta T=40$ K) channel flow for $\Rey=500$, $\mathcal{T}=4$, and $\Pran=7$. The mean flow is along the positive $x$ as marked by arrows in (a) and (b). The colours are the 40$\%$ isosurfaces of the maximum (yellow) and minimum (blue) values of the streamwise perturbations $u_1$. The isosurfaces for other stratification levels ($\Delta T = 20$ K and $60$ K) are qualitatively similar to (b), with 40$\%$ isosurfaces of $u_1$ localised near the hot wall, where viscosity is lower.}
    \label{fig:lin_optimal_u_pert}
\end{figure}

%(\rg{do you mean these people did the comparison for their flows???} - (yes, as Foures was linear 2D, I just decided to remove the whole sentence and just cite Foures. If people are interested they can go and check that paper) 
% \rg{Why is this in green??} \textbf{Needs to be condensed} 

When the direct-adjoint looping is employed at $E_0=10^{-8}$, the optimal perturbation obtained by direct-adjoint looping, and its early time evolution, remain linear. This was remarked upon by \cite{foures2013localization}, and we checked this for stratified flows as well, as will be discussed. By increasing $E_0$, we may attain optimal perturbations which are increasingly nonlinear. We will see below how nonlinear optimal perturbations are very different from the linear, and how this impacts the evolution in a significant manner. 

The optimal perturbations are visualised in this paper as isosurfaces of maximum and minimum streamwise velocity perturbations $u_1$, e.g. as in figure \ref{fig:lin_optimal_u_pert} shown for linear optimal perturbations. In this figure and those to follow, a yellow isosurface is plotted at a certain percentage of the maximum over the channel of that quantity at that time, while a blue isosurface indicates regions where the quantity is at the same percentage of the minimum (usually a negative quantity). 

%   

%The linear optimal perturbations for the unstratified and different degrees of stratification appear in the form of arrays of vortices aligned against the mean flow. 
\begin{figure}
    \centering
    \centerline{\includegraphics[width=0.85\textwidth, height=0.85\textheight,keepaspectratio]{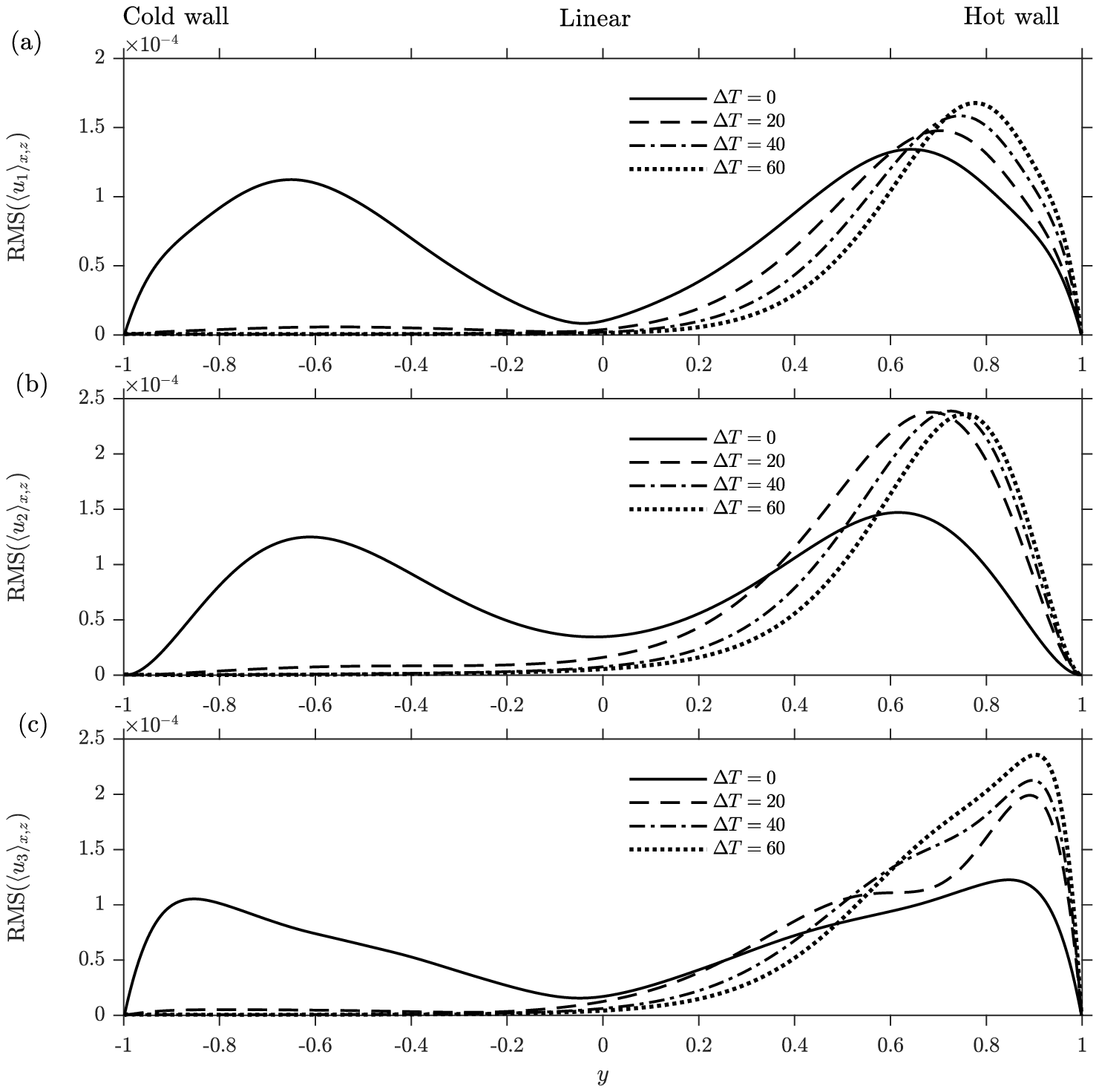}}
    \caption{Wall normal profiles of root mean square (r.m.s., spatially averaged in the $x$ and $z$ directions) of the linear optimal perturbations  ($E_0 = 10^{-8}$). (a) Streamwise velocity perturbations $u_1$, (b) wall-normal velocity perturbations $u_2$, and (c) spanwise velocity perturbations $u_3$ for various wall-temperature differences $\Delta T$ (in K). The solid and the dash-dotted line in (a) correspond to the isosurfaces shown in figure \ref{fig:lin_optimal_u_pert}(a) and (b), respectively.}
     \label{fig:rmsvalues_linear}
\end{figure}

\begin{figure}
    \centering
    \centerline{\includegraphics[width=1.0\textwidth, height=1.0\textheight,keepaspectratio]{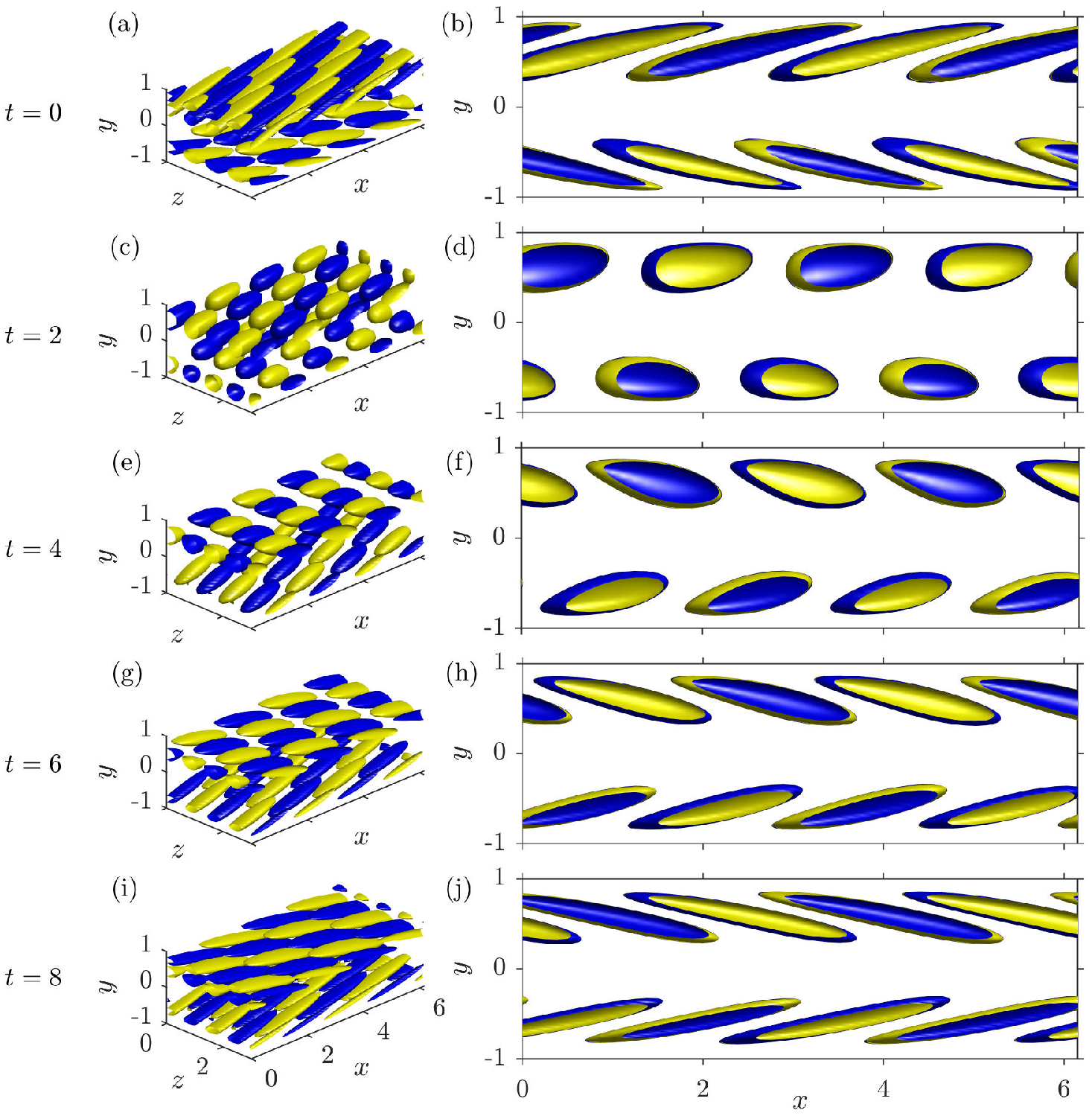}}
    \caption{Evolution of the linear unstratified optimal perturbation shown at two angles, at times (a,b) $t=0$, (c,d) $t=2$, (e,f) $t=\mathcal{T}=4$, (g,h) $t=6$, and (i,j) $t=8$. The structures are initially aligned against the shear, and as time progresses, realign along the shear.}\label{fig:lin_optimal_u_pert_evolution_unstratified}
\end{figure}

The linear optimal perturbation ($E_0=10^{-8}$) consists of an array of streamwise velocity perturbations inclined against the mean flow and shear, on both sides of the channel for the unstratified case (figure \ref{fig:lin_optimal_u_pert}(a)). In the stratified case, similar structures are seen, but all perturbations are remarkably localised close to the hot wall, where viscosity decreases towards the wall, with practically no action on the cold wall (figure \ref{fig:lin_optimal_u_pert}(b)). Such localisation of linear optimal perturbations was also found by \cite{jose2020localisation} using SVD studies on a channel with viscosity-stratification and weak gravity. For our chosen target-time of $\mathcal{T}=4$, we find that the nonmodal energy growth and the shapes of the optimal perturbations are similar whether we optimise for a cost functional with energy growth at the target-time or with time-integrated energy as in equation \eqref{cost_function}. As mentioned, the linear optimal perturbation for maximising energy at a target-time can also be obtained by an SVD of the respective Orr-Sommerfeld and Squire operators for the unstratified \citep{schmid2002stability} and viscosity-stratified \citep{chikkadi2005preventing} cases. The streamwise and spanwise  wavenumbers of the linear optimal perturbation from SVD for $\mathcal{T}=4$ and $\Rey= 500$ are  $k_x\approx2$ and $k_z\approx4$, respectively, for an unstratified channel and $k_x\approx2$ and $k_z\approx5$, respectively, for the viscosity-stratified channel with $\Delta T= 40$ K. Quantized for channel length, we observe from figure \ref{fig:lin_optimal_u_pert} that these wavenumbers can be seen in the linear optimal perturbations obtained from direct-adjoint looping. Besides revealing the localisation of the arrays of vortices near the hot wall due to viscosity stratification, this result is also a validation for our direct-adjoint looping. 
%provides the first source of 

\begin{figure}
    \centering
    \centerline{\includegraphics[width=1.0\textwidth, height=1.0\textheight,keepaspectratio]{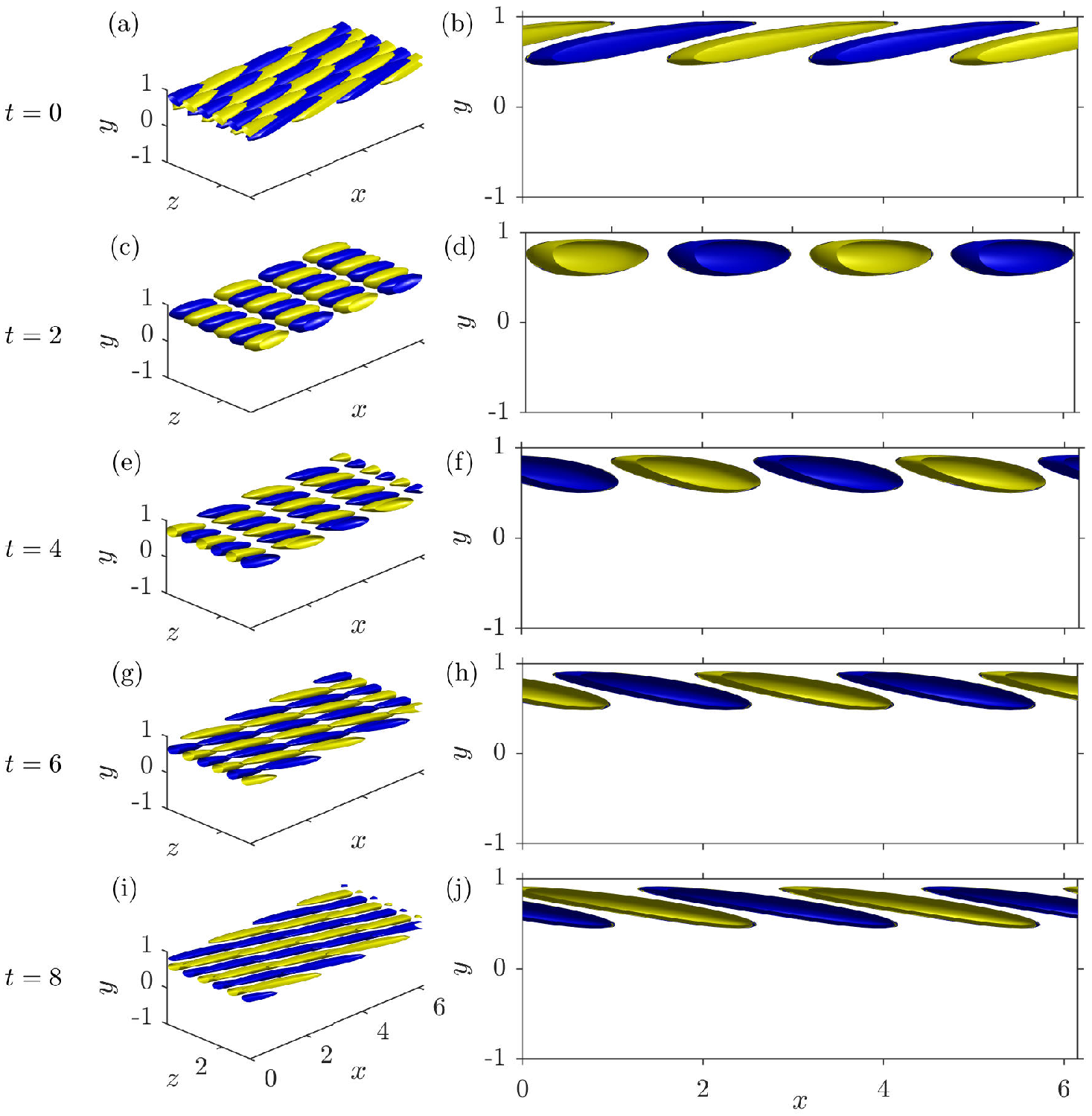}}
    \caption{Evolution of the linear viscosity-stratified optimal perturbation ($\Delta T = 40$ K) shown at two angles. Optimal perturbations are strongly localised on the top (hot) wall unlike in figure \ref{fig:lin_optimal_u_pert_evolution_unstratified}, and the Orr mechanism is in evidence. The times are as in figure \ref{fig:lin_optimal_u_pert_evolution_unstratified}. }\label{fig:lin_optimal_u_pert_evolution_stratified}
\end{figure}

%, at times (a,b) $t=0$, (c,d) $t=2$, (e,f) $t=\mathcal{T}=4$, (g,h) $t=6$, and (i,j) $t=8$. 

The corresponding root mean square (r.m.s.) profiles of velocity perturbations of the linear optimal perturbations are shown in figure \ref{fig:rmsvalues_linear}, where the quantities have been averaged in the streamwise and spanwise directions. There is a significant proportion of initial amplitude in each velocity component, and increase in $\Delta T$ increases the proportion of energy in the spanwise and wall-normal perturbations $u_3$ and $u_2$. The localisation of all perturbations on the hot side of the channel is underlined in this figure. 

\begin{figure}
    \centering
    \centerline{\includegraphics[width=0.65\textwidth, height=0.50\textheight,keepaspectratio]{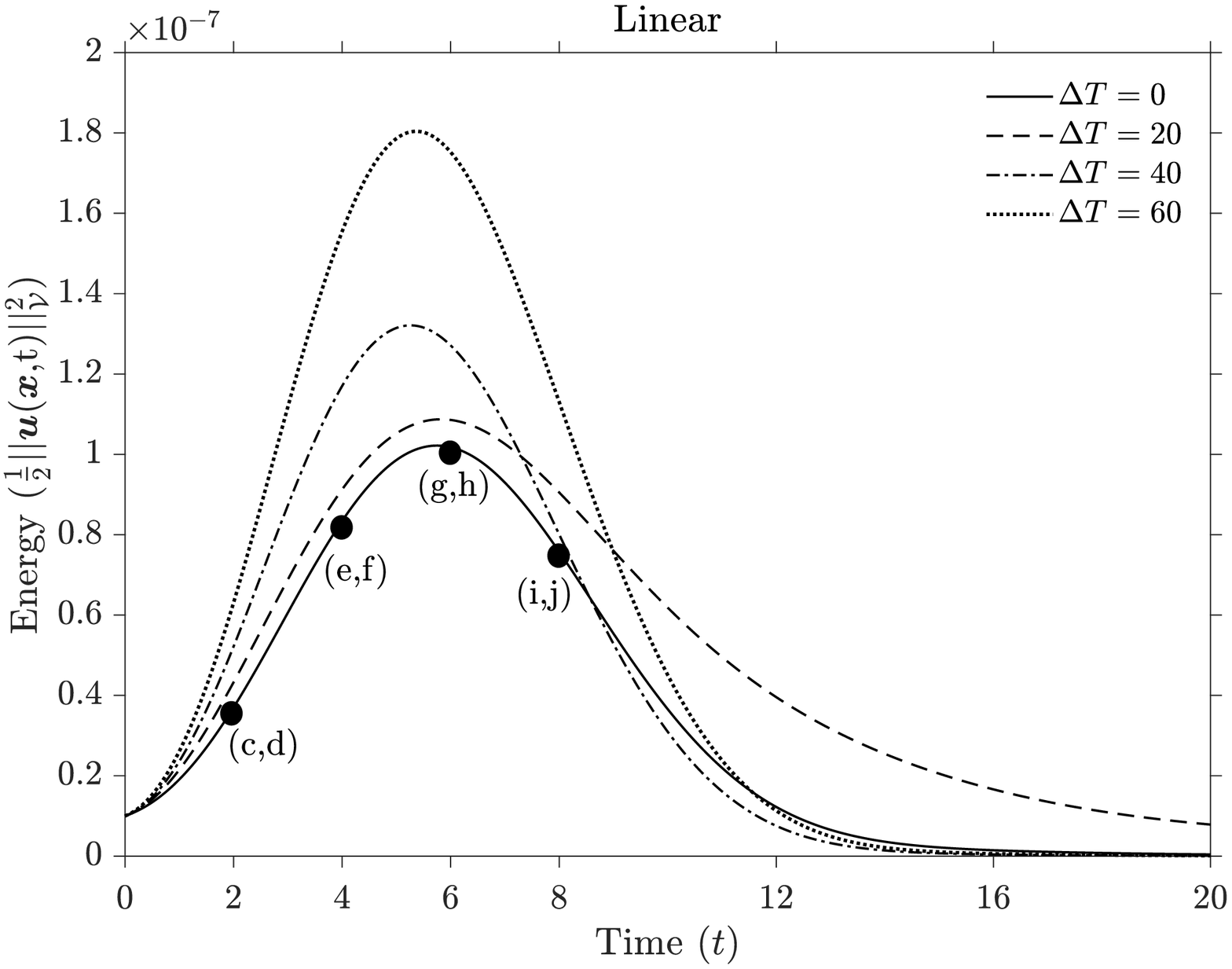}}
    \caption{Energy growth with time of the linear optimal perturbations ($E_0 = 10^{-8}$) for various stratification strengths. The labels at $t=2,4,6,8$ on the solid line correspond to labels in figure \ref{fig:lin_optimal_u_pert_evolution_unstratified}. \label{fig:lin_optimal_u_pert_evolution_energy}}
\end{figure}

%\textcolor{red}{should the energy have double angle brackets i.e. $\frac{1}{2}\langle \langle\mathbf{u}(\mathbf{x},t)\cdot\mathbf{u}(\mathbf{x},t)\rangle \rangle$} 
The time evolution, obtained by solving the direct equations initialised with the linear optimal perturbation, suggests the reason for its shape. For both the unstratified and stratified cases, shown in figures \ref{fig:lin_optimal_u_pert_evolution_unstratified} and \ref{fig:lin_optimal_u_pert_evolution_stratified} respectively, velocity perturbations are initially tilted against the mean shear, and as time progresses, lean into the shear as they stretch. This is the well known, and probably oldest to be described, linear growth mechanism, the Orr mechanism \citep{orr1907stability}, where the titling and the subsequent energy growth is driven by the base, or laminar, shear. In stratified laminar flow, the magnitude of shear is larger near the less viscous wall, which for liquids is the hot wall (figure \ref{plot:profiles}(b)). So, the Orr mechanism is much more efficient near the hot wall. It follows that for a given $E_0$, better growth can be achieved by placing perturbations in the high gradient region, which explains the localisation of the initial velocity perturbations in stratified flow (figure \ref{fig:lin_optimal_u_pert}(b) and \ref{fig:rmsvalues_linear}). The evolution of the optimal perturbations result in algebraic energy growth of disturbances for short duration of time which eventually decays as shown in figure \ref{fig:lin_optimal_u_pert_evolution_energy}. For the linear optimal perturbations, the energy growth for stratified flow is larger than for unstratified flow, but this conclusion will not be the same for the nonlinear optimal perturbations, as we shall see.

\begin{figure}
    \centering
    \centerline{\includegraphics[width=1.0\textwidth, height=1.0\textheight,keepaspectratio]{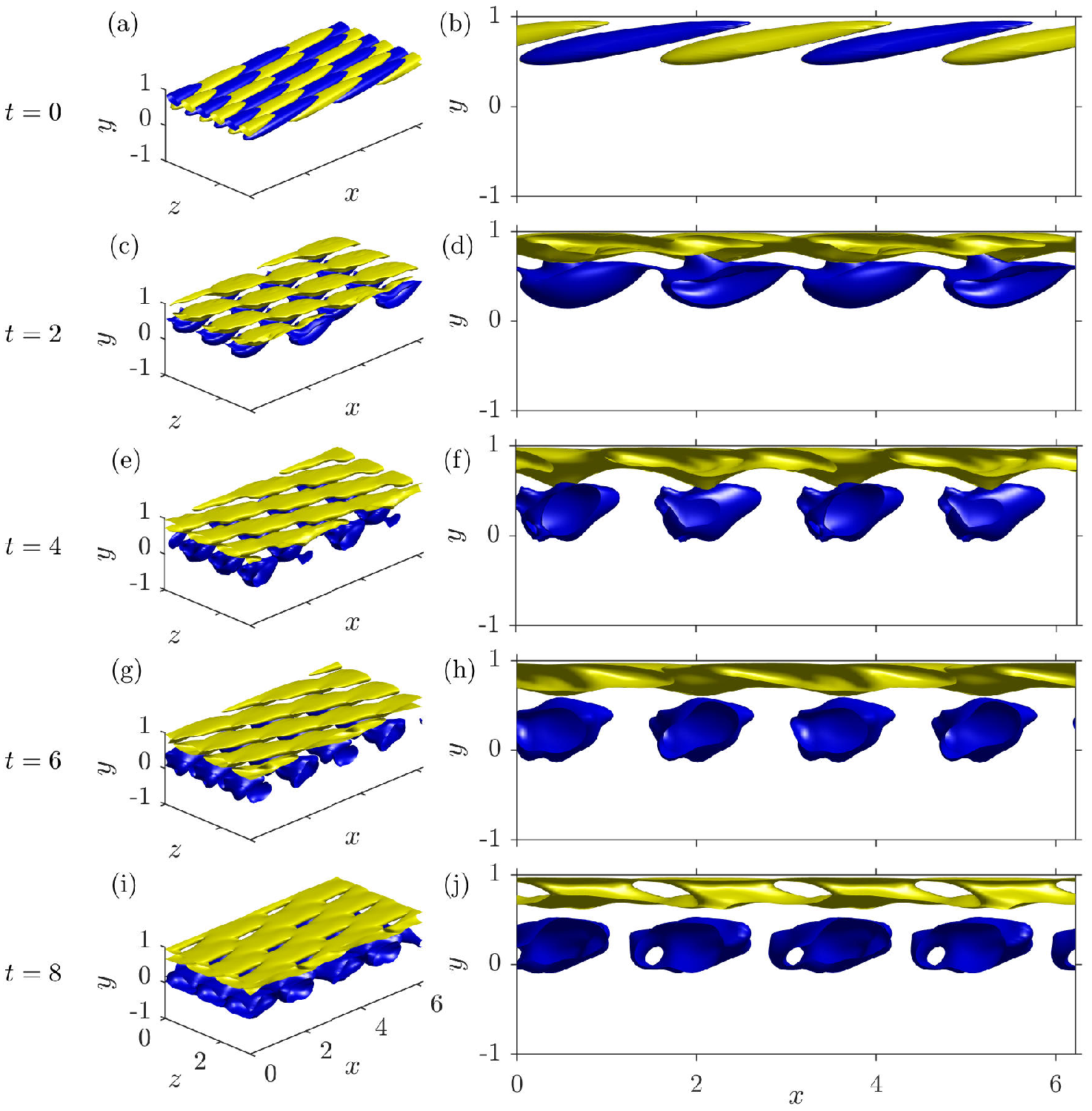}}
    \caption{Evolution of the linear viscosity-stratified optimal perturbation ($\Delta T = 40$ K) scaled to nonlinear initial energy $E_0 = 10^{-2}$, shown at two angles. The lift-up mechanism is in evidence. The times are as in figure \ref{fig:lin_optimal_u_pert_evolution_unstratified}.  \label{fig:linearinnonlinearupertevol}}
\end{figure}
We thus find that the Orr mechanism is the dominant linear growth mechanism for small energy levels in this short target-time window. The other well-known linear growth mechanism, the lift-up mechanism \citep{brandt2014lift}, is not observed in the evolution of linear optimal perturbation at small $E_0$. Before we study nonlinear optimal perturbations, it is instructive to study what would happen if the linear optimal perturbation was in large enough amplitude to trigger nonlinearities. To this end, we rescale the energy of the linear optimal perturbations to a higher initial energy, $E_0=10^{-2}$, while maintaining the shape of the initial conditions corresponding to the case shown in figure \ref{fig:lin_optimal_u_pert}(b) for $\Delta T = 40$ K. The evolution of the streamwise velocity perturbations for this case is shown in figure \ref{fig:linearinnonlinearupertevol}. The low momentum fluid is transferred away from the walls, displaying features of the classical lift-up mechanism \citep{brandt2014lift} driven by streamwise vortices (not shown). Comparing figures \ref{fig:lin_optimal_u_pert_evolution_stratified} and \ref{fig:linearinnonlinearupertevol} we see that the nonlinear evolution of the linear optimal perturbation is very different from the linear evolution of the linear optimal perturbation. The non-linear evolution of the linear optimal perturbation for the unstratified case (not shown) also shows a lift-up type mechanism in operation, albeit at both walls, and is symmetric about $y=0$. The physical mechanism for energy growth at small energy levels ($E_0 = 10^{-8}$) is thus the Orr mechanism and that at high energy levels ($E_0 = 10^{-2}$) is indicative of the lift-up mechanism. As we discuss below, in particular for stratified flow, the linear optimal perturbations are not the most efficient way to extract energy from the mean flow into the perturbations for higher energy levels.

\subsection {The non-linear optimal perturbation and its evolution} 

\begin{figure}
    \centering
    \centerline{\includegraphics[width=1.0\textwidth, height=1.00\textheight,keepaspectratio]{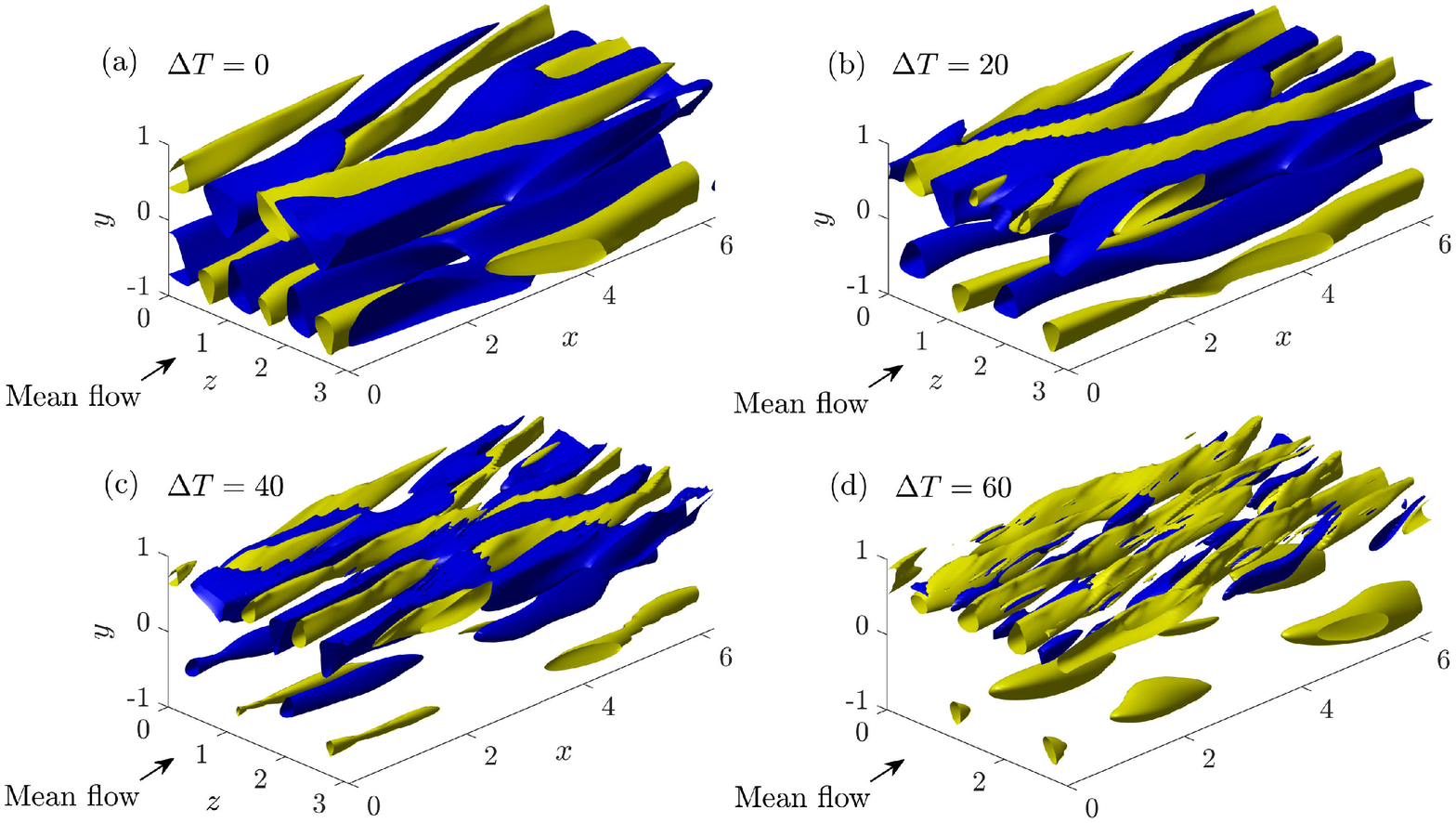}}
    \caption{40$\%$ isosurfaces of the maximum (yellow) and minimum (blue) values of the streamwise perturbations $u_1$ of the nonlinear optimal perturbation ($E_0 = 10^{-2}$) with (a) $\Delta T = 0$ (unstratified); and of the viscosity-stratified nonlinear optimal perturbation with (b) $\Delta T = 20$ K and (c) $\Delta T = 40$ K. (d) 20$\%$ isosurfaces of the maximum (yellow) and minimum (blue) $u_1$ for the viscosity-stratified nonlinear optimal perturbation with $\Delta T = 60$ K. A slightly lower isosurface value had to be shown in (d) for better visualisation. }
    \label{fig:optimalNonLin}
\end{figure}

%The nonmodal energy growth curves corresponding to the nonlinear optimal perturbations in (a-d) are shown in figure \ref{fig:energyNonlin}.
%\textcolor{red}{We should improve these figures to show noisy structures, especially Delta T=60 plot is misleading.}

\begin{figure}
    \centering
    \centerline{\includegraphics[width=0.85\textwidth, height=0.85\textheight,keepaspectratio]{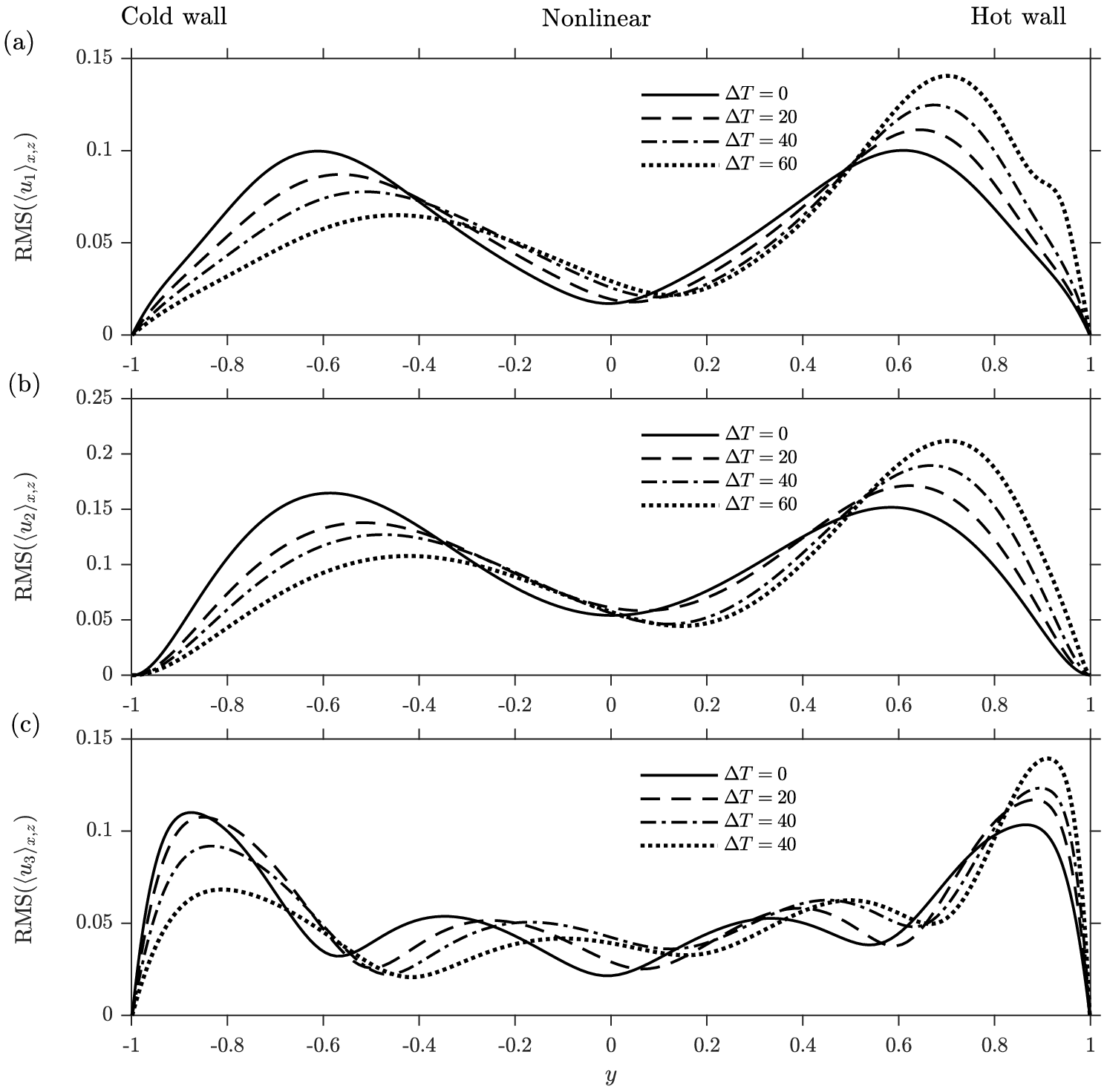}}
    \caption{Wall normal profiles of (a) streamwise velocity fluctuations $u_1$, (b) wall-normal velocity fluctuations $u_2$, and (c) spanwise velocity fluctuations $u_3$, averaged across the $x$ and $z$ coordinates, of the nonlinear optimal perturbations ($E_0 = 10^{-2}$) for various stratification strengths.}
    \label{fig:rmsvaluesNonlinear}
\end{figure}

The perturbation leading to the maximum energy growth for the highest $E_0$ of $10^{-2}$ considered here, is referred to as the nonlinear optimal perturbation. Isosurfaces of the nonlinear optimal streamwise velocity perturbation (figure \ref{fig:optimalNonLin}) and  streamwise-spanwise averaged r.m.s. wall-normal nonlinear optimal velocity profiles (figure \ref{fig:rmsvaluesNonlinear}) show some localisation towards the hot wall due to viscosity stratification. But remarkably, unlike in the linear case (figures \ref{fig:lin_optimal_u_pert} and \ref{fig:rmsvalues_linear}), there is  significant perturbation energy on both walls of the channel for the stratified nonlinear optimal perturbation. Figure  \ref{fig:rmsvaluesNonlinear}, in stark contrast to figure \ref{fig:rmsvalues_linear}, makes it clear that the asymmetry between the two sides of the channel is small for the nonlinear optimal perturbation, whereas in the linear optimal perturbation, energetic structures were absent in the bottom half of the channel. But in the nonlinear optimal perturbation too, the asymmetry increases with increasing stratification, with more structures at the hot wall. The streamwise velocity perturbations are now arranged in a series of elongated (mainly in the flow direction, but with a spanwise inclination) high and low momentum zones near the walls. Increasing the stratification level makes the population near the cold wall smaller (but not insignificant).  From figure \ref{fig:rmsvaluesNonlinear} we observe a significant contribution to the initial perturbation kinetic energy from all three components of velocity.

%\textcolor{red}{should the energy have double angle brackets i.e. $\frac{1}{2}\langle \langle\mathbf{u}(\mathbf{x},t)\cdot\mathbf{u}(\mathbf{x},t)\rangle \rangle$}
\begin{figure}
    \centering
    \centerline{\includegraphics[width=1.0\textwidth, height=1.0\textheight,keepaspectratio]{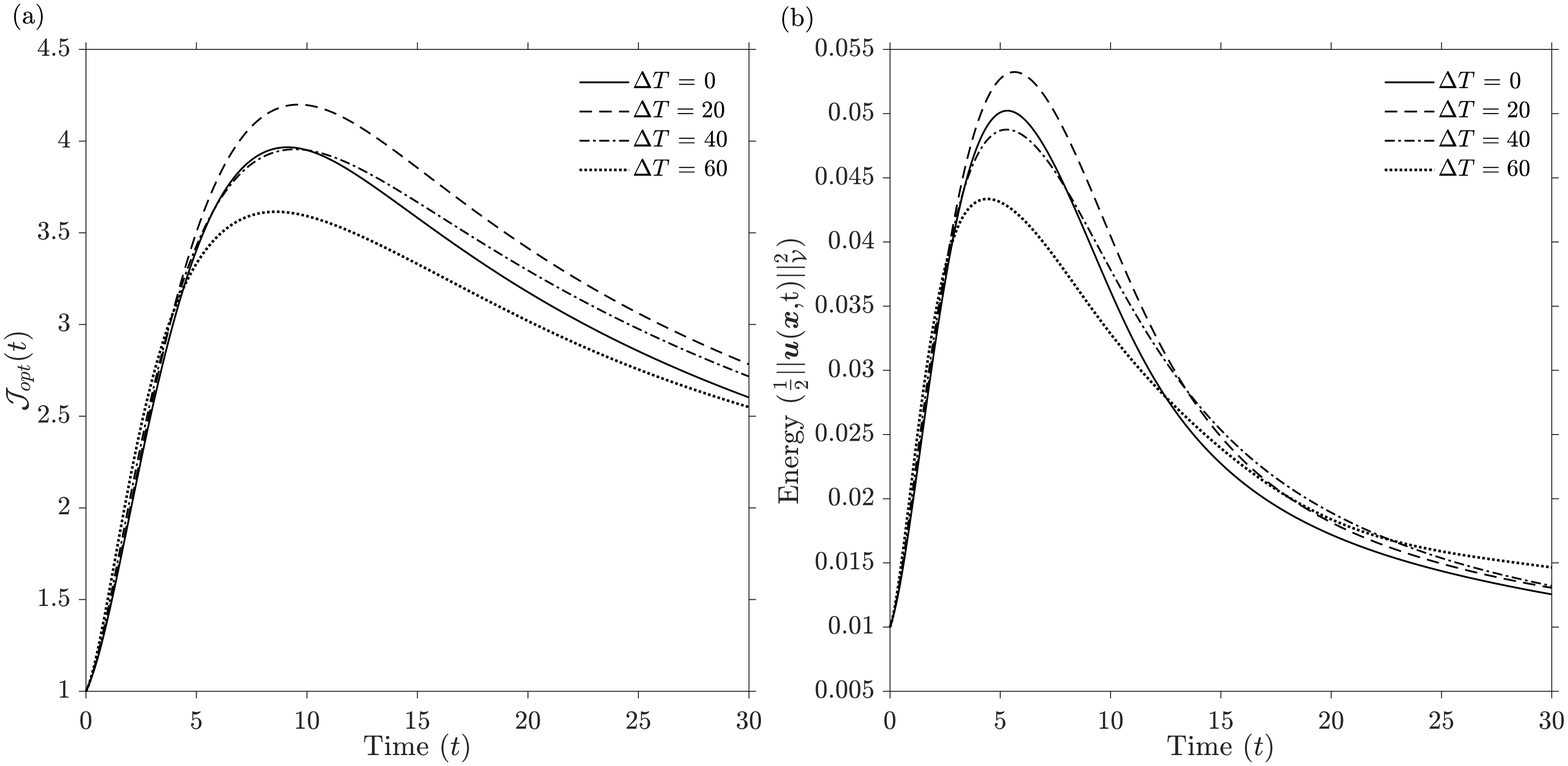}}
    \caption{Time evolution of (a) the cost functional $\mathcal{J}_{opt}(t)$ as in equation \eqref{eq:jopt} and of (b) energy, of the nonlinear optimal perturbations for various stratification strengths. The time of optimisation is $\mathcal{T}$=4 for all. }
    \label{fig:energyNonlin}
\end{figure}
\begin{figure}
    \centering
    \centerline{\includegraphics[width=0.60\textwidth, height=0.60\textheight,keepaspectratio]{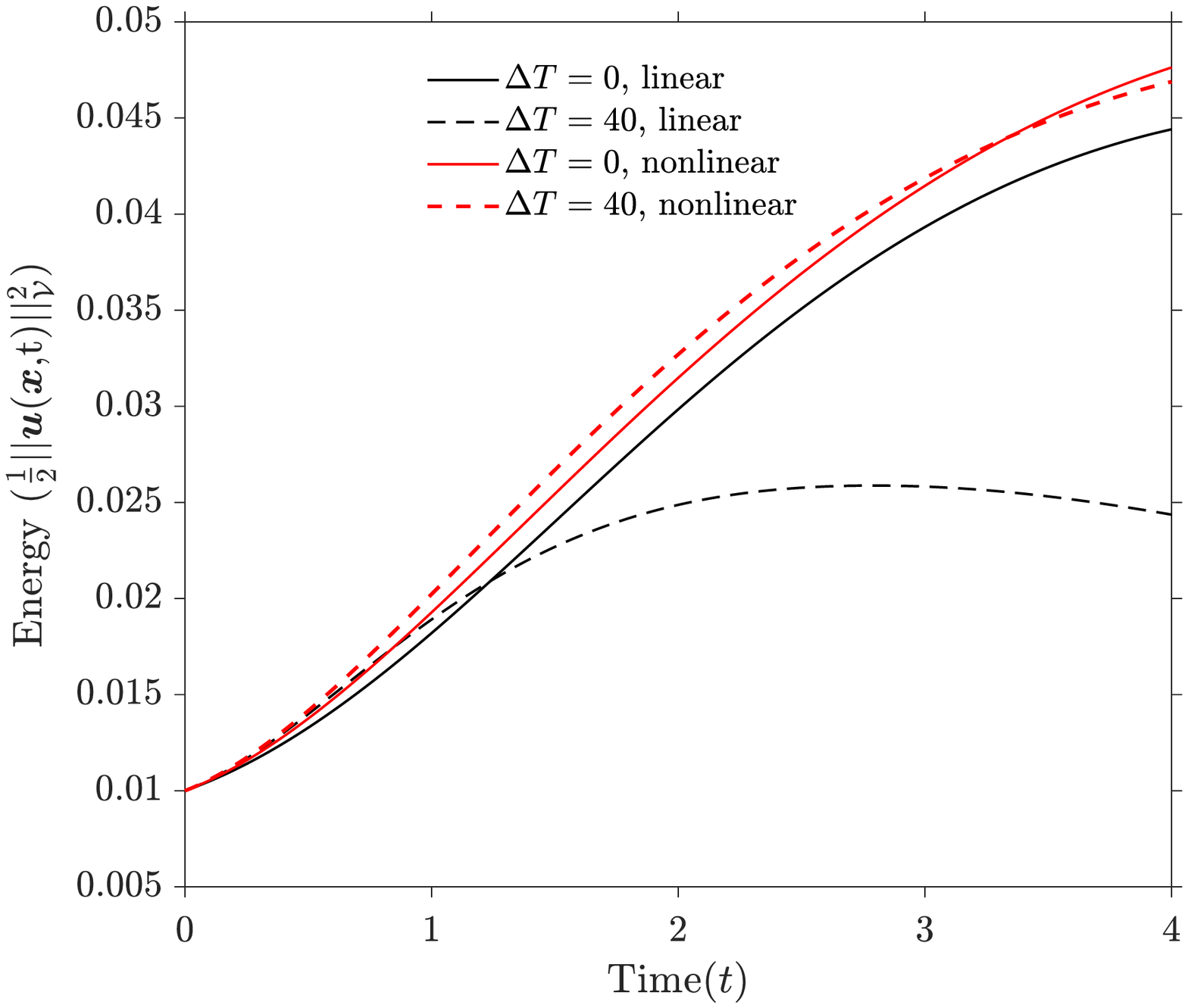}}
    \caption{Evolution of energy of the linear (black lines) and nonlinear (red lines) optimal perturbations when evolved with the modified Navier-Stokes equation with $E_0=10^{-2}$. Solid lines are for the unstratified cases ($\Delta T = 0$) while the dotted lines are for stratified cases with $\Delta T = 40$ K. Other stratification levels (not shown), show similar behaviour to $\Delta T = 40$ K. 
    \label{fig:linearinnonlinearEnergytime}}
\end{figure}
%\textcolor{red}{should the energy have double angle brackets i.e. $\frac{1}{2}\langle \langle\mathbf{u}(\mathbf{x},t)\cdot\mathbf{u}(\mathbf{x},t)\rangle \rangle$}
The energy-time graphs corresponding to the evolution of the nonlinear optimal perturbations in figures \ref{fig:optimalNonLin} and \ref{fig:rmsvaluesNonlinear} are shown in figure \ref{fig:energyNonlin} for various stratification levels. Figure \ref{fig:energyNonlin}(a) shows the cost functional $\mathcal{J}_{opt}(t)$ of the optimal perturbation which is the quantity that we optimised for, while figure \ref{fig:energyNonlin}(b) shows the volume-averaged kinetic energy as a function of time. Growth is algebraic in the nonlinear regime as well, and perturbations decay soon after the target-time of optimisation. Unlike in the linear evolution of the linear optimal perturbation, there is no qualitative difference between the growth in the unstratified case and those at various levels of stratification. We are now in a position to compare the evolution, by the modified Navier-Stokes equations, of the linear and the nonlinear optimal perturbations, both starting from the same initial energy of $E_0=10^{-2}$, in figure \ref{fig:linearinnonlinearEnergytime}. We may first satisfy ourselves of the higher energy growth in the evolution of the non-linear optimal perturbation as compared to the linear optimal perturbation, consistent with the definition of the nonlinear optimal perturbation. For the unstratified flow, for a short target-time, it turns out that the linear and nonlinear optimal perturbations show similar growth, though the linear is of course lower, whereas the linear optimal perturbation shows a much lower growth in the stratified flow than the nonlinear optimal perturbation (comparing the dashed black line to the dashed red one in figure \ref{fig:linearinnonlinearEnergytime}). This is consistent with the cold wall becoming more prominent in the evolution of the nonlinear optimal perturbation, as we shall discuss below. It is also worth mentioning that on comparison with figure \ref{fig:lin_optimal_u_pert_evolution_energy} we see that the growth of energy of the linear optimal perturbation, as a ratio of the initial energy, is significantly lower with nonlinear evolution, for initial conditions differing only in amplitude. However, the absolute value of perturbation energy always remains larger than the linear case since the initial perturbation was large. When the initial perturbation is large, the available energy from the laminar flow becomes a limiting factor, which could result in the lower growth, as a ratio.

We now discuss how stratification changes the mechanism of subcritical disturbance growth and how nonlinear optimal perturbations are fundamentally different from linear optimal perturbations in this regard. Initially proposed by \citet{hamilton1995regeneration} and \citet{waleffe1997self} and summarised by \citet{brandt2014lift}, the regeneration/self-sustaining cycle of wall turbulence involves three steps, (i) lift-up, i.e., transportation of low (high) momentum fluid away from (towards) the wall by streamwise vortices, to form streamwise independent streaks of low (high) momentum away from (near) the wall, (ii) break down of these by inflectional secondary instability to acquire streamwise dependence and (iii) regeneration of elongated vortices by nonlinear interactions between oblique modes. These arguments were initially made with the linear optimal perturbation in mind. Through a direct-adjoint looping optimisation methodology,  \citet{cherubini2011minimal} for a boundary layer and \citet{cherubini2013nonlinear} for a Couette flow showed that it is much more efficient for the lift-up to be driven by streamwise modulated vortices in the first place. The nonlinear optimal perturbation inherently contains such streamwise modulation. This is referred to as the modified lift-up mechanism, as it can bypass the stage of secondary (streak) instability en-route to transition to turbulence. 
%, at times (a,b) $t=0$ , (c,d) $t=2$, (e,f) $t=4$, (g,h) $t=6$, (i,j) $t=8$. 
\begin{figure}
    \centering
    \centerline{\includegraphics[width=1.0\textwidth, height=1.0\textheight,keepaspectratio]{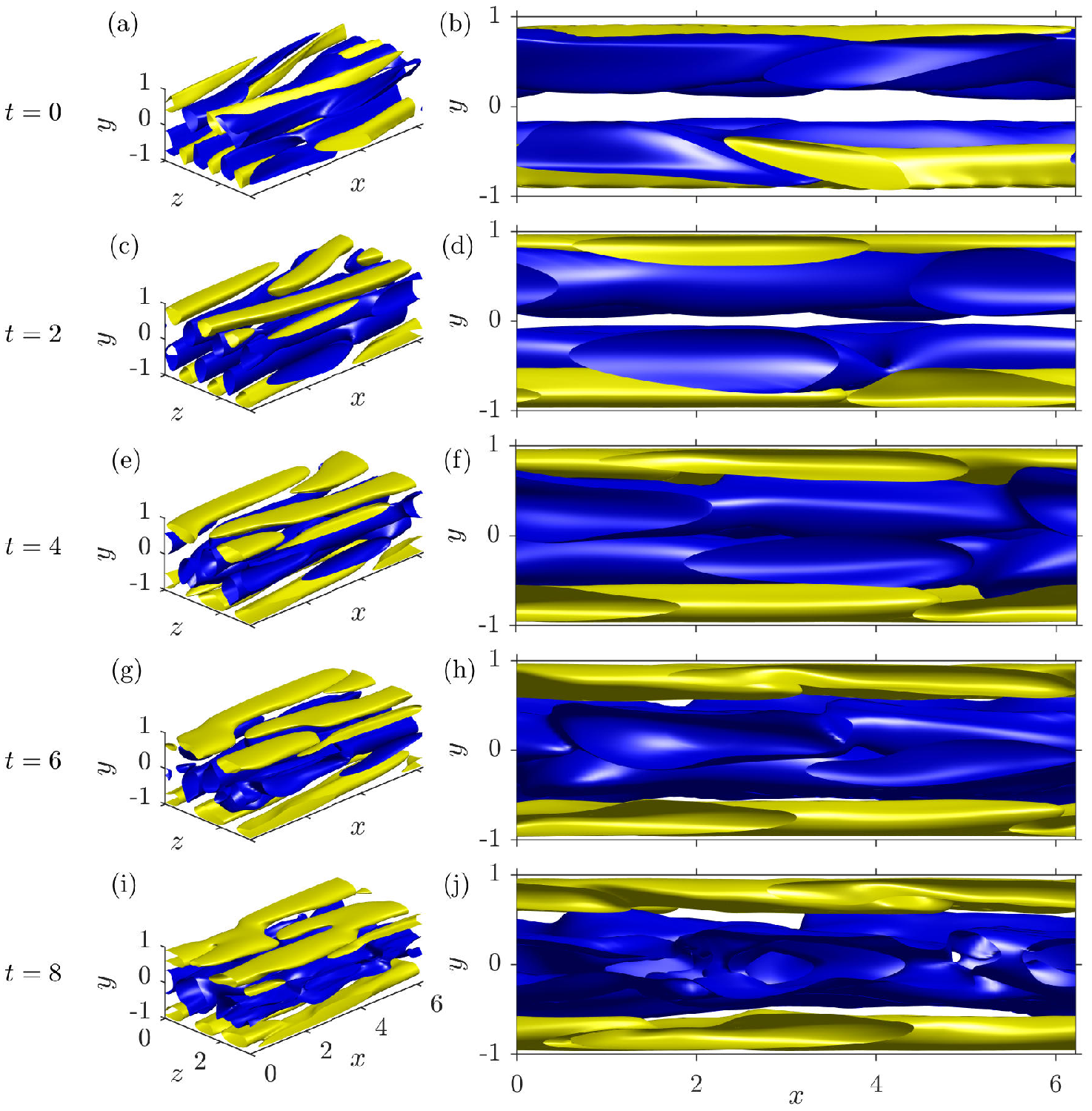}}
    \caption{Evolution of the nonlinear unstratified optimal perturbation with $E_0=10^{-2}$, shown at two angles. The times are as in figure \ref{fig:lin_optimal_u_pert_evolution_unstratified}. \label{fig:nonLinOptEvol}}
\end{figure}
We detect similar optimal perturbation structures here for a channel flow, both in the unstratified and stratified cases. Their evolution in time by the modified Navier-Stokes equation is shown in figure \ref{fig:nonLinOptEvol} for the unstratified case. A modified lift-up mechanism similar to \cite{cherubini2011minimal} and \cite{cherubini2013nonlinear} is seen to be in operation, where low momentum fluid is lifted off the wall and high momentum fluid is brought closer to the wall by streamwise modulated vortices (vortices not shown). This translates into the algebraic growth of perturbation kinetic energy seen in figure \ref{fig:energyNonlin}. 

\begin{figure}
    \centering
    \centerline{\includegraphics[width=1.0\textwidth, height=1.0\textheight,keepaspectratio]{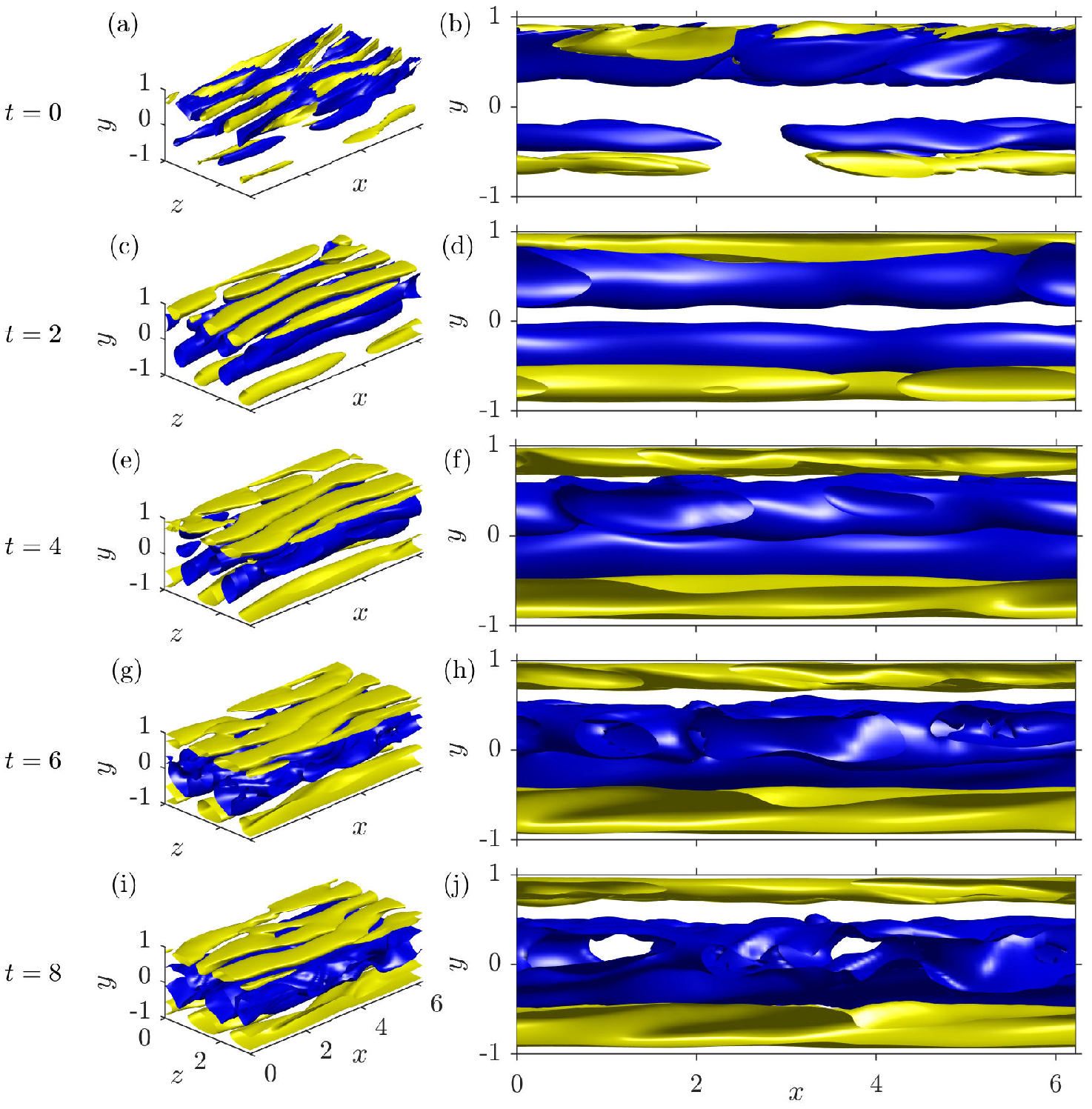}}
    \caption{Evolution of the nonlinear viscosity-stratified optimal perturbation ($\Delta T = 40$ K) with $E_0=10^{-2}$, shown at two angles. The times are as in figure \ref{fig:lin_optimal_u_pert_evolution_unstratified}. \label{fig:nonlin_strat_evolution}}
\end{figure}
\begin{figure}
    \centering
    \centerline{\includegraphics[width=1.0\textwidth, height=1.0\textheight,keepaspectratio]{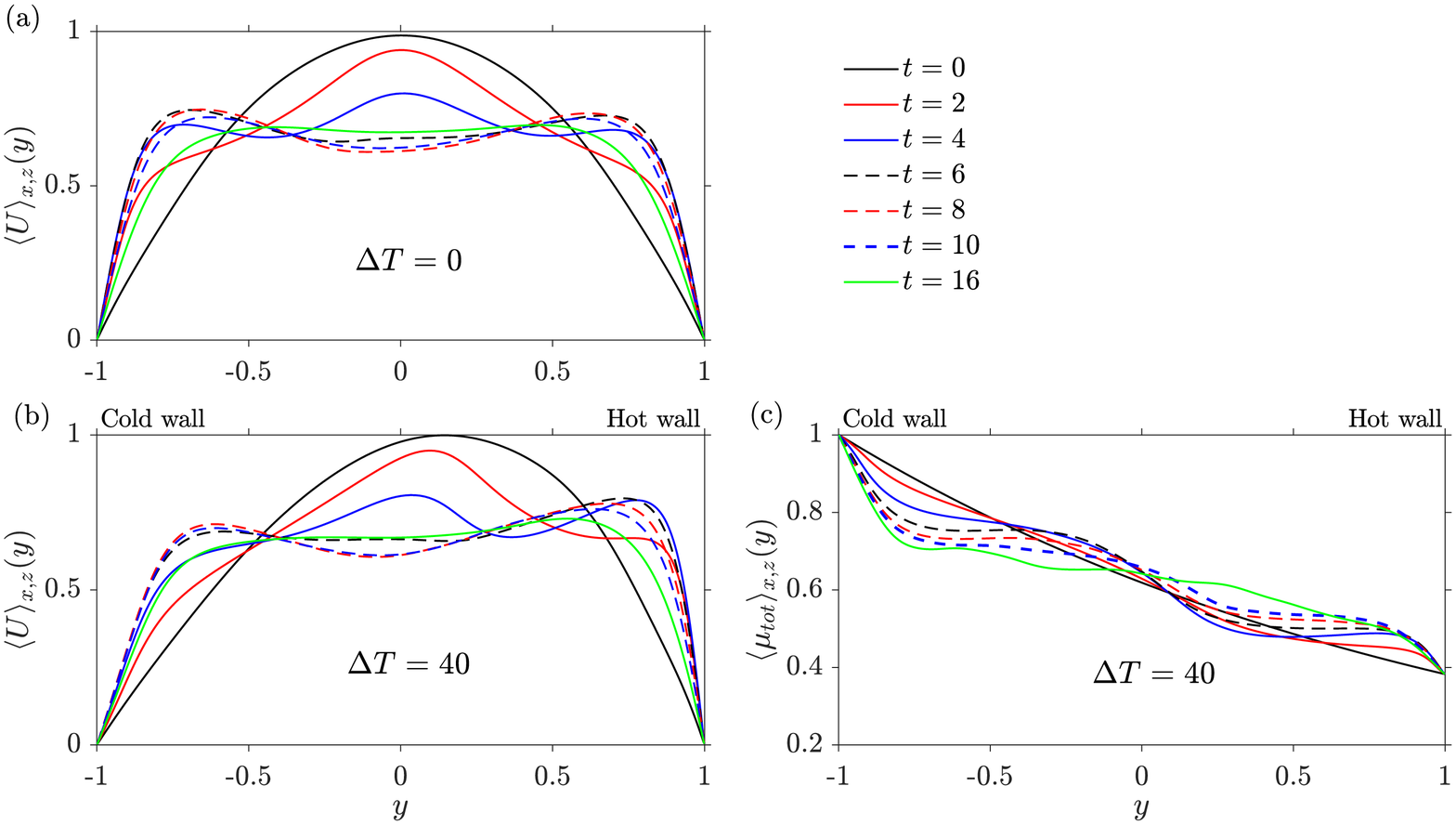}}
    \caption{The total streamwise velocity at various times averaged across the $x$ and $z$ coordinates for (a) the nonlinear unstratified optimal perturbation and (b) the nonlinear viscosity-stratified optimal perturbation for $\Delta T=40$ K. (c) The evolution of the total viscosity profile for the flow corresponding to (b). Solid black lines in each for $t=0$, solid red for $t=2$, solid blue for $t=4$, dashed black for $t=6$, dashed red for $t=8$, dashed blue for $t=10$, and solid green for $t=16$.\label{fig:inflection_points}}
\end{figure}

%(\textcolor{blue}{earlier it was $\partial{U_1}/\partial{y}$. As we do not have anything like $U_1$, I have made it $U$}\textcolor{red}{in the equations we refer to U_i, but in the profiles we refer to U, I think its probably too much pain to backtrack and change things consistently, everyone will probably understand with U or U_1, so we leave it as you have written}
The evolution of the nonlinear optimal perturbation in stratified flow is shown in figure \ref{fig:nonlin_strat_evolution}. As mentioned earlier, the inception of an inflection point due to lift-up may be expected to be more efficient near the less viscous wall as the wall-normal velocity gradient is larger, and lift-up is usually associated with $u_2 \partial{U}/\partial{y}$  \citep{cherubini2011minimal}. Consistent with this, we have a larger population of optimal perturbation structures near the less viscous wall, as seen in figure \ref{fig:optimalNonLin}(a) and (b). Since mean shear is smaller at the cold wall, its lift-up capability is lower, and therefore it may be argued that it is structures which are already a little away from the cold wall, which can grow better on the cold side. This is borne out by the optimal perturbation structures seen in the bottom half of figure \ref{fig:nonlin_strat_evolution}(b). 
An interesting feature of the evolution of the viscosity stratified nonlinear optimal perturbation, which distinguishes it from the unstratified case as well as from the evolution of the linear optimal perturbation, is that as time progresses action at the cold wall is increasingly significant, and the high-speed structures at the hot wall shrink in wall-normal extent. The evolution of perturbations at the cold wall is strong enough to create points of inflection in the $x-z$ averaged velocity profiles, and this will be discussed with the aid of figure \ref{fig:inflection_points}. We shall refer to a "strengthening (weakening)" of inflectional profiles when the profile becomes more (less) strongly wavy in the wall-normal direction. In panel (a) of this figure, we see that the unstratified flow progresses steadily towards inflection, maintains this up to about $t=10$ and become less inflectional thereafter. The profiles are symmetric. In fact the perturbations in all cases decay at long times, given low Reynolds number and small target-time of optimisation employed here. In figure \ref{fig:inflection_points}(b) the evolution of the profile in the stratified case is shown. There is a strengthening of the inflectional profile at early times at both walls, with the hot wall being more inflectional. After about $t=4$, the profile becomes weaker near the hot wall and more strongly inflectional than before on the cold side, before eventually weakening at long time. The corresponding profiles of the $x-z$ averaged total viscosity are shown in figure \ref{fig:inflection_points}(c). Upon comparing with the laminar viscosity profile, up to a time of about $10$, on the colder side, we see that higher viscosity fluid from the cold wall has been lifted up towards the centreline and lower viscosity fluid from the central portion of the channel has been carried towards the wall. A similar exchange is visible on the hot side of the channel as well, but with opposite signs of viscosity change. At the long time of $t=16$, we see a mixing of fluids. 
Below, we address the question: if lift-up is more efficient at the less viscous wall, why do perturbations grow near the other wall?

%\rg{Please call the y-axis $\mu_{tot}$ in this figure (c)} 
\begin{figure}
    \centering
    \centerline{\includegraphics[width=1.0\textwidth, height=1.0\textheight,keepaspectratio]{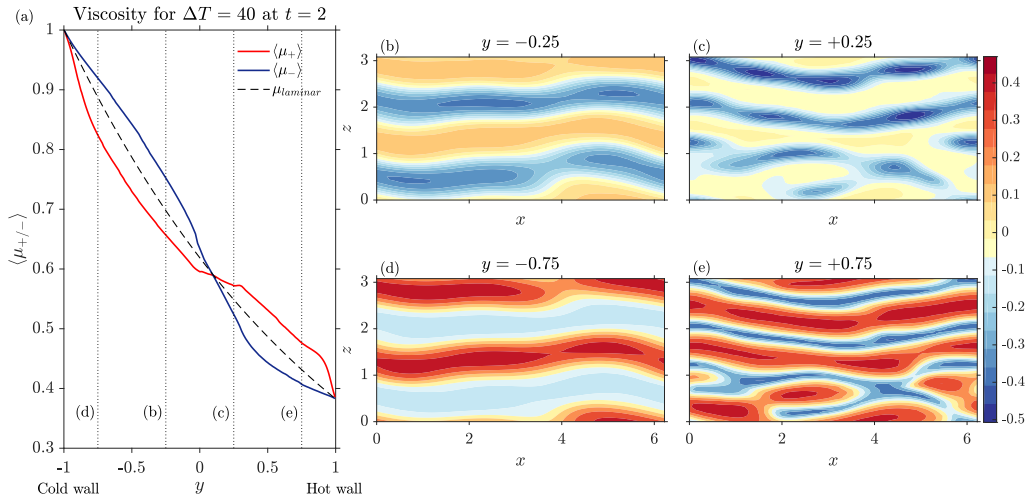}}
    \caption{Flow driven by the nonlinear viscosity-stratified optimal perturbation at $\Delta T = 40$ K and time $t=2$. (a) Viscosity profiles averaged across the $x$ and $z$ coordinates for positive streamwise velocity perturbation $u_1 > \epsilon$ in red, and for $u_1 < -\epsilon$ in blue. The laminar viscosity profile is shown as a dashed black line. The four vertical black dotted lines with labels denote the $y$ locations of the plots in (b-e). Instantaneous streamwise velocity perturbations $u_1$ are shown in the $x$-$z$ plane at $y$ locations (b) -0.25, (c) 0.25, (d) -0.75, and (e) 0.75. Refer to figure \ref{fig:nonlin_strat_evolution}(c) for a 3D view of isosurfaces of $u_1$ at this time and the red solid line in figure \ref{fig:inflection_points}(b) for the total $U(y)$ averaged in $x$ and $z$ at this time. \label{fig:visc_slices}}
\end{figure}
\begin{figure}
    \centering
    \centerline{\includegraphics[width=1.0\textwidth, height=1.0\textheight,keepaspectratio]{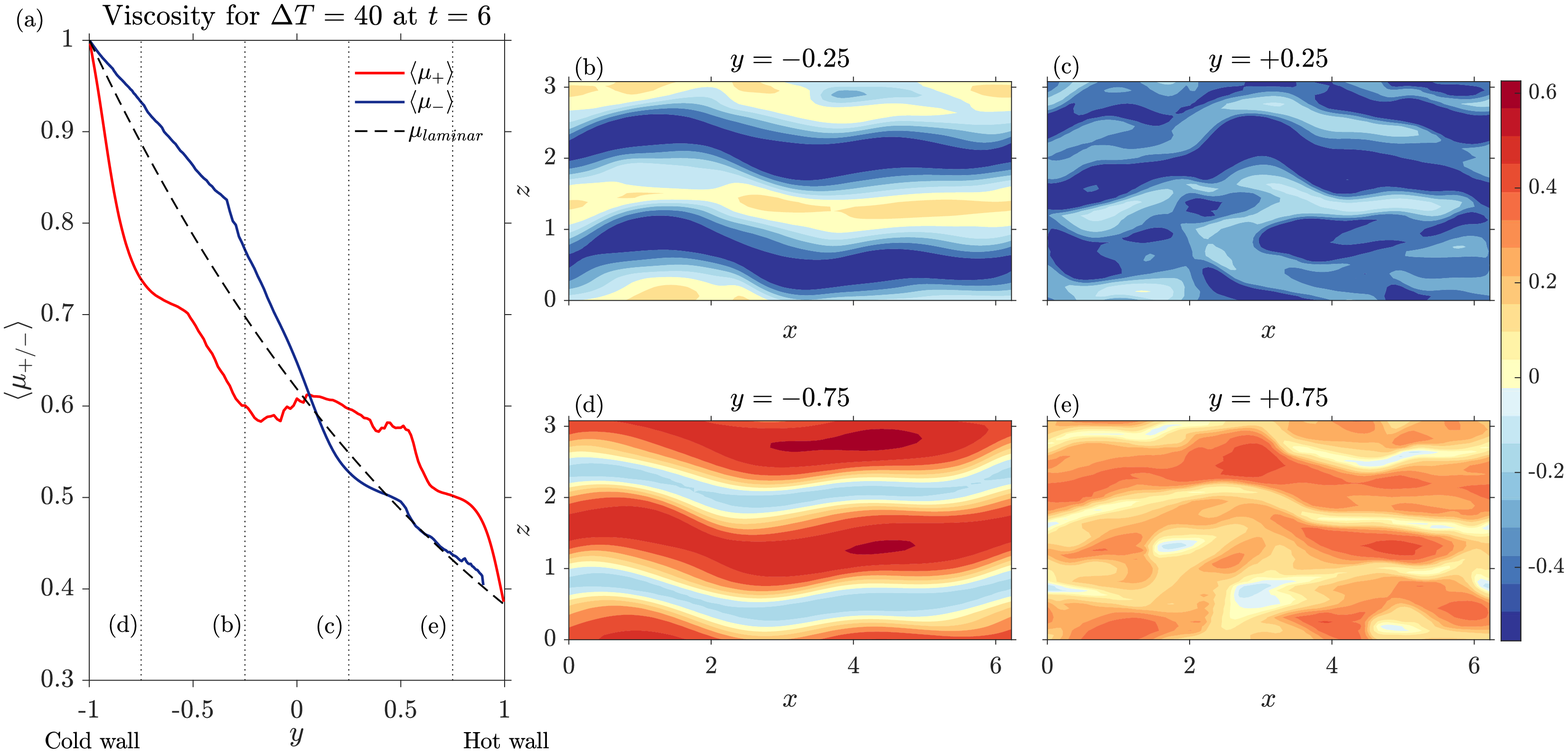}}
    \caption{Same as figure \ref{fig:visc_slices} but for time $t=6$. Refer to figure \ref{fig:nonlin_strat_evolution}(g) for a 3D view of isosurfaces of $u_1$ at this time and the black dashed line in figure \ref{fig:inflection_points}(b) for the total $U(y)$ averaged in $x$ and $z$ at this time.
    \label{fig:visc_slices2}}
\end{figure}
% \rg{Please call the y-axis $\langle{\mu_{+/-}}\rangle$ in this figure and following one too. In the figure you can replace $u>...$ by $\langle{\mu_+}\rangle$} 
After the inception of lift-up, near the hotter wall, less viscous fluid of low momentum is brought away from the wall to the vicinity of more viscous and high momentum fluid, and the opposite happens on the colder wall. Thus the low (high) momentum streaks near the hotter wall are composed of less (more) viscous fluid, but those near the colder wall are composed of more (less) viscous fluid than the local laminar values. This is evident from the conditionally averaged viscosity profiles in figures \ref{fig:visc_slices}(a) and \ref{fig:visc_slices2}(a) at time $t=2$ and $6$ respectively. Here, at each $y$ location, the viscosity $\langle{\mu_+}\rangle$ is averaged over all positive $u_1$ cells over the $x-z$ plane, and $\langle{\mu_-}\rangle$ is the viscosity averaged over negative $u_1$ cells in the same plane. A small cut-off $\epsilon=0.001$ in the velocity has been used for this averaging, and it has been checked that the profiles are insensitive to the exact choice of $\epsilon$. These plots establish that on the colder side of the channel, low speed regions are correlated with reduced viscosity, and high speed regions correlate with elevated viscosity, with the opposite correlations on the hotter side. The instantaneous streamwise velocity perturbations in four different $x-z$ planes are shown in figures \ref{fig:visc_slices} and \ref{fig:visc_slices2} in panels (b) to (e). The spanwise widths and spacing of the low and high speed streaks is significantly larger on the cold side than on the hot side. Secondly the streaks persist up to $t=6$ on the cold side, whereas on the hot side the structure is practically lost by this time. 
A physical argument for the relative persistence near the colder wall is as follows. Consider that the streamwise pressure gradient is similar across the span of the channel. High speed streaks of low viscosity alternating with low speed flow of higher viscosity would be maintained by this pressure gradient. On the other side, i.e., at the hot wall, a higher viscosity fluid of higher forward speed would tend to slow down, and a higher viscosity fluid of lower speed to speed up, in response to a similar streamwise pressure gradient. The greater persistence of streaks on the colder side is thus a consequence of the basic asymmetry in the mechanics of the lift up on the two sides. 

In figure \ref{fig:optimalNonLin}(c) we had seen that the spanwise variation of the velocity perturbations had different apparent wavenumbers on the two walls. The slices in figures \ref{fig:visc_slices} and \ref{fig:visc_slices2} clarify this to be a sinuous variation. Such variation is known to be responsible for the ultimate breakdown of streaks \citep{waleffe2009exact}. A study at higher Reynolds number and longer target-time could reveal this. Besides, the inflectional instability, discussed below for the stratified case, extracts energy from the streaks \citep{waleffe2009exact} allowing further energy growth beyond the lift-up. Further studies at higher Reynolds numbers and longer target-times will be needed to explore these mechanisms in viscosity-stratified flows.

\begin{figure}
    \centering
    \centerline{\includegraphics[width=1\textwidth, height=1\textheight,keepaspectratio]{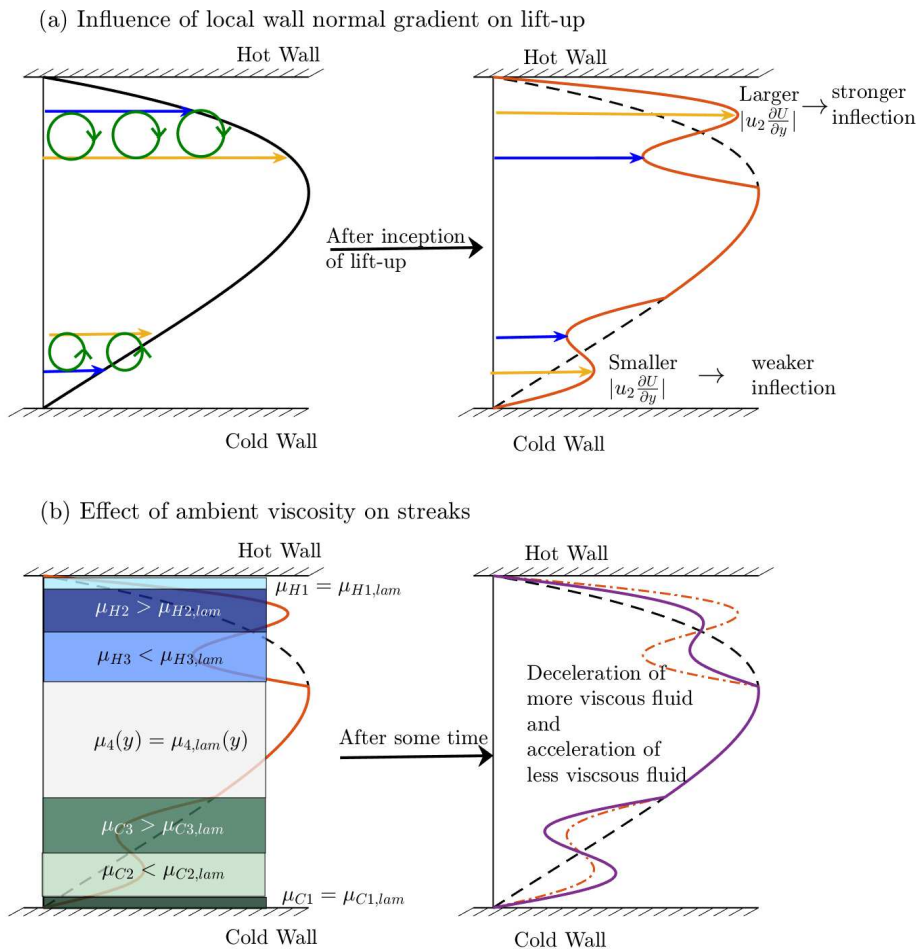}}
    \caption{ Schematic of the lift-up mechanism influenced by viscosity stratification: (a) inception of the inflection in the velocity profile is stronger near the less viscous wall as it has larger wall-normal velocity gradient, (b) persistence of the inflection created is greater near the cold/more-viscous wall because the streak $C2$ of high momentum can sustain higher wall-normal gradients of velocity than before, whereas the streak of low momentum, $C3$ has higher viscosity velocity gradients within it will be lowered. The opposite happens on the other wall where high momentum fluid $H2$ has higher viscosity and low momentum fluid $H3$ has lower viscosity than the local laminar value. The dashed line represents the undisturbed laminar profile, the dash-dotted line and the solid lines are representations of early and later times respectively.}
    \label{fig:SchematicLiftUp}
\end{figure}
The observations in figures \ref{fig:inflection_points} to \ref{fig:visc_slices2} enable us to schematically illustrate the lift-up process in stratified flow, in figure \ref{fig:SchematicLiftUp}. Panel (a)  shows stronger inception of inflection near the hot wall at early time. On the left of panel (b) we sketch how this lift-up results in exchange of viscosities. On the right of this panel, we see how this exchange of viscosities results in strengthening of the inflectional profile on the cold wall and weakening on the hot wall. Wherever viscosity is higher than the surrounding laminar flow, gradients are lowered and wherever it is lower, gradients are relatively increased. Thus, the low viscosity streak at the cold wall brings with it higher velocity gradients, leading to stronger lift-up. The persistence of high momentum and low viscosity streaks, combined with stronger inflection in the velocity profile near the colder wall is consistent with the observations of previous DNS studies \citep{zonta2012modulation,lee2013effect} concerned with turbulence in stratified flow. In boundary layer flow \citep{lee2013effect} heating the flat plate and hence making fluid less viscous in the vicinity leads to suppression of turbulence and for channel flow \citep{zonta2012modulation} turbulence is suppressed on the hot/less viscous wall and enhanced on the cold/more viscous wall. \citet{cherubini2011minimal} and \citet{cherubini2013nonlinear} highlight the importance of the Orr and lift-up mechanisms, both linear mechanisms, in the creation of sub-critical transition through  minimal seeds of turbulence transition (obtained by optimizing over much larger target-times as compared to what we study in this paper). Recently, \citet{Vavaliaris2020} also reported the dominance of these mechanisms in the initial stages of sub-critical turbulence in a boundary layer. We have shown how viscosity stratification in a channel acts to modify these mechanisms. For the short target-time ($\mathcal{T}=4$) optimal perturbations at the relatively small Reynolds number ($\Rey=500$) that we have studied, the interaction required for non-linear regeneration of the streaks and hence completing the regeneration cycle en-route to transition \citep{waleffe2009exact} is absent. But the primary role of viscosity stratification in the initial stages of the nonlinear nonmodal process has been revealed.

\subsection{Effect of Prandtl number}

\begin{figure}
    \centering
    \centerline{\includegraphics[width=1.1\textwidth, height=1.1\textheight,keepaspectratio]{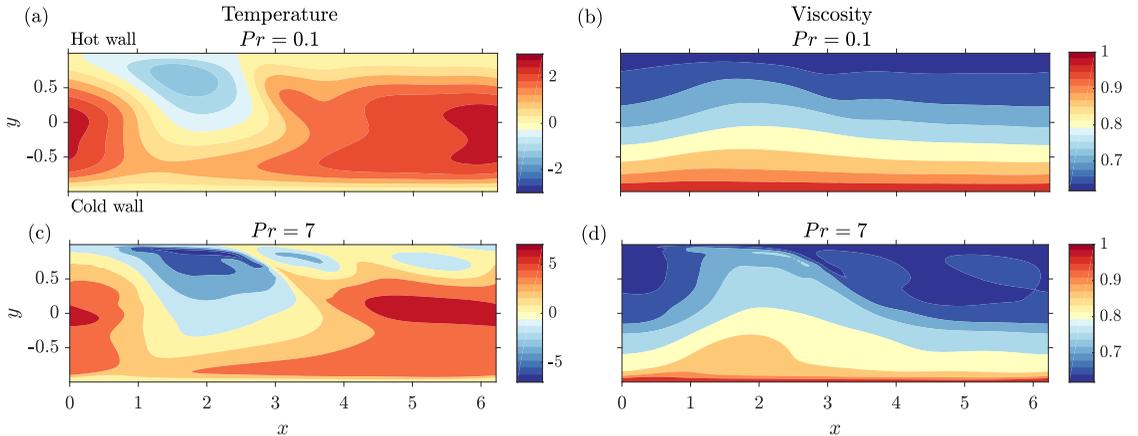}}
    \caption{Temperature perturbations at target-time at $z = \upi/2$ for (a) $\Pran=0.1$ and (c) $\Pran=7$ when started with the corresponding nonlinear viscosity-stratified optimal perturbation ($\Delta T = 20$ K). The corresponding viscosity contours at same time are in (b) and (d). Note the presence of higher gradients in temperature and viscosity in (c) and (d), respectively. The colour bars in (a) and (c) are different.}
    \label{fig:temp_evolve_slice}
\end{figure}
We performed simulations at three Prandtl numbers: 
$\Pran=0.1, \ 7$ and $5000$, for $\Delta T=20$ K. Our lowest Peclet number, i.e., the product of the Reynolds and the Prandtl numbers, is $50$, which is too large for diffusion of the temperature perturbations to qualitatively change the behaviour over our simulation times. We confirm this in our simulations. Slices of temperature and viscosity perturbations are shown in figure \ref{fig:temp_evolve_slice} for two values of $\Pran$, when evolved with the corresponding nonlinear optimal perturbation up to the target-time. We see that diffusion effects are greater at the lower Prandtl number, so viscosity variations persist better at the higher $\Pran$, while we find very similar structures and their evolution (not shown) at all Prandtl numbers. However, in studies over longer target-times, of the process of transition to turbulence, we expect the Prandtl number to play an important role.

\section{Conclusions}

In this study we have derived, for the first time to our knowledge, the adjoint modified Navier-Stokes equations for a viscosity-stratified flow. We have developed a numerical solver for the direct and adjoint equations in an iterative loop in three dimensions, to find optimal initial conditions in the linear limit as well as for a finite initial perturbation energy. We have shown that viscosity stratification brings important modifications to the operation of the lift-up mechanism in the early stages of disturbance growth. Initially stronger lift up is set up at the hot (less viscous) wall due to the higher mean velocity gradient, but the lift-up at the cold (more viscous) wall increases in strength later, while that at the hotter wall weakens. Significantly, at the colder wall, high-speed streaks are more persistent, of larger spanwise extent, and give rise to a strengthening of the inflectional profile. We have presented physical arguments for these observations. Thus the action shifts from the hotter wall to the colder wall as time progresses. Most of the features we observe in the evolution of the nonlinear optimal perturbation are completely missed in the linear study. A linear optimal perturbation of small amplitude will only display the Orr mechanism and not the lift-up. At higher amplitudes, lift up will be seen, but only at the hot wall. In fact no perturbations are ever seen near the cold wall with the linear optimal perturbation. 

This work suggests several directions for future research. A starting point for understanding the role of viscosity stratification in the transition to turbulence will be the study of nonlinear optimal perturbations over long target-times. The question of the minimal seed for triggering turbulence could be of importance in this context.  \cite{vermach2018optimal} made an interesting finding that the initial condition which produces the most efficient mixing could be quite different from that which gives the highest energy growth. Recognising that most flows where questions about mixing are relevant are also stratified in viscosity indicates this as an area of study. We expect the effect of Prandtl number to be pronounced in flows with a sharp stratification, e.g., the flow of miscible fluids of different viscosity, and also at long times in continuously stratified flows, and this bears investigation. We have neglected gravity in this study but most flows with a composition to temperature variation are subject to buoyancy effects. This combination will make for interesting study. Finally, given the number of industrial applications for which viscosity stratification is important, a variety of experimental studies are called for. We hope that this first work on the effects of viscosity stratification in nonlinear optimal perturbation growth will give impetus to such studies.\\
\\

\noindent \textbf{Declaration of Interests}\\ The authors report no conflict of interest.

\section*{Acknowledgements}
George Batchelor, and stories about him, have been an inspiration to us over the years, and it is a pleasure to dedicate this paper to his memory. The authors would like to thank Sharath Jose and Vishal Vasan for helpful discussions during the development of the solver and on nonmodal analysis. RT and RG acknowledge support, under project no. 12-R\&D-TFR-5.10-1100 of the Department of Atomic Energy and the Ocean Mixing and Monsoon program of the Ministry of Earth Sciences, of the Government of India. Simulations were performed in the high performance computing facility of ICTS, Bengaluru.

\appendix 
\section{More on the numerical method and its validation}
The $x$ and $z$ directions in figure \ref{plot:channel} are the homogeneous directions and we define periodic boundary conditions in these, except in the pressure. We impose no-slip boundary conditions at the walls at $y=\pm L_y$. For all the analysis, we use a nondimensional size of $L_x = 2\upi$, $L_y = 2$, and $L_z = \upi$ in which the streamwise extent is smaller than in \citet{vermach2018optimal}, while the spanwise extent is the same. Our channel size is the same as in the 2D study of \citet{foures2013localization}.

The gradients in the velocity and temperature fields are higher with nonlinear initial energy ($E_0 = 10^{-2}$) when compared to linear initial energy ($E_0 = 10^{-8}$). In $x$, $y$, and $z$, we use 100 $\times$ 209 $\times$ 50 grid points for $E_0 = 10^{-2}$, while we use  50 $\times$ 209 $\times$ 25 grid points for $E_0 = 10^{-8}$. The makes the grid spacing in the $x$ and $z$ directions equal, at 0.06 for $E_0 = 10^{-2}$ and at 0.12 for $E_0 = 10^{-8}$. For a nonlinear stratified case with $\Delta T = 20$ and $E_0 = 10^{-2}$, the maximum difference between $E(\mathcal{T}) =  \frac{1}{2}\langle \langle\mathbf{u}(\mathbf{x},t)\cdot\mathbf{u}(\mathbf{x},t)\rangle \rangle$ with a grid of 100 $\times$ 209 $\times$ 50 versus a finer grid of 128 $\times$ 209 $\times$ 64 (this produces grid spacing the same as that of \citet{vermach2018optimal}) is $O(10^{-5})$. 
%\sout{$G(\mathcal{T}) = ||\boldsymbol{u}(\boldsymbol{x},\mathcal{T})||_{\mathcal{V}}^{2}$}
%\rg{We need to say how many iterations are typical for our case, and whether stratified converges slower than unstratified} 

The wall normal or the $y$ direction is discretised into a staggered combination of base and fractional grids (see e.g. \citet{bewley2012numerical}). This puts a fractional grid point at the mid-location of two base grid points. The streamwise and spanwise velocities, pressure, and temperature are defined in the fractional grid and the wall-normal velocity in the base grid. A hyperbolic tangent function 
\begin{equation}
    y_j = \tanh\left( k \left[ \frac{2(j-1)}{N_Y} - 1 \right]    \right), \quad j=1,2, \dots N_Y,
    \label{stretch}
\end{equation}
where $N_Y$ is the number of grids in the wall-normal direction, is used to cluster both the fractional and the base grid points towards the walls. The value of $k$ in equation \ref{stretch} has been kept constant at 1.5 and it creates a $y$-grid with $\Delta y_{max}$ = 0.0159 (near the centre of the channel) and $\Delta y_{min}$ = 0.0029 (near both the walls). The wall normal spatial derivatives are computed using second order central finite difference method. The spatial derivatives in $x$ and $z$ are calculated using Fast-Fourier transform and we truncate the Fourier series using the 2/3-rule to prevent aliasing (see e.g. \citet{canuto2007spectral}). We employ a time stepping algorithm which is a combination of an explicit method (Runge-Kutta-Wray) for the nonlinear (convective) terms and an implicit method (Crank-Nicolson) for the linear (viscous) terms and the wall-normal derivatives. 

To converge to the optimal perturbation, we start with a guess of the same. After we have completed one iteration by going forward and backward with the direct and adjoint equations, we need to update the guess for the next iteration. We employ the update-by-rotation technique of \cite{foures2013localization} which involves a line search after the adjoint leg of the direct-adjoint loop is completed. This first amounts to updating the direct variables at $t=0$ after the $n^{th}$ iteration by an angle $\zeta$, between 0$^\circ$ and 360$^\circ$, as \begin{equation}\label{eq:zetaeqn}u_i^{n+1}=u_i^{n}\cos{\zeta}+f_i(\mathbf{u}^{n},\mathbf{v}^{n})\sin{\zeta},\end{equation} where the exact functional form of $f_i$ in terms of the direct, $\mathbf{u}^{n}$, and adjoint, $\mathbf{v}^{n}$, velocities is provided in \cite{foures2013localization}. The next step involves evolving the direct equations with $u_{i}^{n+1}$ for each update angle $\zeta$ and noting the cost functional. The update angle that provides the largest cost functional is chosen as our desired update angle $\zeta_{max}$. The iteration continues with the updated guess $u_i^{n+1}=u_i^{n}\cos{\zeta_{max}}+f_i(\mathbf{u}^{n},\mathbf{v}^{n})\sin{\zeta_{max}}$.
%This update angle along with the adjoint and direct variables at $t=0$ from the previous iteration provides the updated initial conditions for the direct variables according to the formulae provided in \cite{foures2013localization} and . \textcolor{red}{The update of the velocity, $u_i$ for the $(n+1)^{th}$ iteration, for example is given by,}
%\sout{\rg{how does one update a variable by an angle, please explain}} 
We have tried various step sizes for the angle search and, after a few iterations, found the direct-adjoint loop to ultimately converge to a zero angle, i.e., where the direct variables from the previous iteration are the optimal perturbations ($u_i^{n+1}=u_i^{n}$) within the tolerance of the step size. As expected, the number of iterations required to converge to zero angle is inversely proportional to the angle step size.  As mentioned in section \ref{dal}, at this point, we find the residual as defined in other studies \citep{vermach2018optimal} to be of $O(10^{-3} - 10^{-4})$. For the geometry considered in this paper, the direct-adjoint loop optimises for $O(10^6)$ variables. It takes $O(100)$ iterations to converge to the nonlinear optimal perturbation, with $E_0 = 10^{-2}$, for the unstratified as well as the stratified cases, similar to the number in \citet{vermach2018optimal}, while it takes less than 50 iterations to converge to the linear optimal perturbation ($E_0 = 10^{-8}$, both unstratified and stratified) when starting from random initial conditions. Imposing an optimal perturbation of a similar parameter set as our initial guess allows much faster convergence to the same optimal perturbation as before, akin to method of continuation. The ease of convergence to the linear optimal perturbation has also been reported in the studies cited in the text. 
%\sout{As mentioned in section \ref{dal}, the rotation technique of} \cite{foures2013localization} \sout{converges when the required update angle is zero.}  

Only for the validation purpose with \citet{vermach2018optimal}, we use a longer channel with $L_x=4\upi$. With our direct-adjoint looping solver, for  $\mathcal{T}=2$, we achieve a $G(\mathcal{T})$ of over 93$\%$ of that of \citet{vermach2018optimal}, which incidentally is the only existing study to our knowledge on nonlinear optimal perturbation in unstratified three dimensional channel flow. The isosurfaces of the streamwise velocity that we obtain for the nonlinear optimal perturbation with $E_0=10^{-2}$, starting from random initial conditions, closely agree with those of \citet{vermach2018optimal}. In particular, both computations result in optimal perturbations which are elongated in the streamwise direction, with streamwise wavelength far larger than can be accommodated in the channel of $4\upi$ length, spanwise wavenumber of $k_z=6$, and very similar levels of obliqueness. A more perfect numerical agreement is not to be expected as these optimal perturbations are known to be numerically delicate \citep{foures2013localization} and might vary with varying resolution of the optimiser. Nonlinearity makes the problem one of nonconvex optimisation and it is not clear how to arrive at the global maximum, or how the existence of more than one local maximum can be resolved \citep{kerswell2018nonlinear}. Still, we have achieved a very good match with a completely independently developed solver for the nonlinear optimal perturbation in the unstratified channel. A check we performed to make sure we have converged to a local optimum is to start from different initial conditions of random noise of the desired amplitude, as well as an optimal perturbation from a different parameter set, to ensure that the optimal perturbations we arrive at are the same within the finite-difference errors ($O(10^{-4})$). For the viscosity-stratified case we ran the direct-adjoint loop at very low initial energy of $E_0=10^{-8}$ to get the linear optimal perturbation by maximising $E(\mathcal{T})$, compared it to the linear viscosity-stratified optimal perturbation obtained from SVD, and obtained excellent agreement as already mentioned in section 3.1.

% \sout{\rg{Is the updat angle identically zero???} \textcolor{red}{we tried various delta Angle and it seemed to go to exact zero for all}} 

%The $y$ direction is discretised with a stretched grid to cluster grid points near the walls and is solved using finite difference methods. The two periodic directions $x$ and $z$ are solved spectrally.  We solve for these three-dimensional perturbations with the modified Navier-Stokes equations for viscosity-stratified flows as in \eqref{eq:Direct_momentum}, subject to incompressibility as in \eqref{eq:Direct_incompressibility}, and the temperature equation as in \eqref{eq:Direct_scalar} using implicit-explicit time marching scheme.

\bibliographystyle{jfm}
%\normalem
\bibliography{references.bib}

\end{document}